\documentclass[sigconf, nonacm]{acmart}

\usepackage{algorithm}
\usepackage[noend]{algorithmic}
\usepackage[outdir=./]{epstopdf}

\newcommand\vldbdoi{XX.XX/XXX.XX}
\newcommand\vldbpages{XXX-XXX}
\newcommand\vldbvolume{14}
\newcommand\vldbissue{1}
\newcommand\vldbyear{2020}
\newcommand\vldbauthors{\authors}
\newcommand\vldbtitle{\shorttitle} 
\newcommand\vldbavailabilityurl{http://vldb.org/pvldb/format_vol14.html}
\newcommand\vldbpagestyle{plain} 
\begin{document}

\title{Estimating Spread of Contact-Based Contagions in a Population Through Sub-Sampling}

\author{Sepanta Zeighami}
\affiliation{%
  \institution{USC}
}
\email{zeighami@usc.edu}

\author{Cyrus Shahabi}
\affiliation{%
  \institution{USC}
}
\email{shahabi@usc.edu}

\author{John Krumm}
\affiliation{%
  \institution{Microsoft Research}
}
\email{jckrumm@microsoft.com}

\setlength{\textfloatsep}{0pt}
\setlength{\floatsep}{0pt}
\setlength{\intextsep}{0pt}
\setlength{\dbltextfloatsep}{0pt}
\setlength{\dblfloatsep}{0pt}
\setlength{\abovecaptionskip}{0pt}
\setlength{\belowcaptionskip}{0pt}
\begin{abstract}
Physical contacts result in the spread of various phenomena such as viruses, gossips, ideas, packages and marketing pamphlets across a population. The spread depends on how people move and co-locate with each other, or their mobility patterns. How far such phenomena spread has significance for both policy making and personal decision making, e.g., studying the spread of COVID-19 under different interventions strategies such as wearing a mask. In practice, mobility patterns of an entire population is never available, and we usually have access to location data of a subset of individuals. In this paper, we formalize and study the problem of estimating the spread of a phenomena in a population, given that we only have access to sub-samples of location visits of some individuals in the population. We show that simple solutions such as estimating the spread in the sub-sample and scaling it to the population, or more sophisticated solutions that rely on modeling location visits of individuals do not perform well in practice, the former because it ignores contacts between unobserved individuals and sampled ones and the latter because it yields inaccurate modeling of co-locations. Instead, we directly model the co-locations between the individuals. We introduce PollSpreader and PollSusceptible, two novel approaches that model the co-locations between individuals using a \textit{contact network}, and infer the properties of the contact network using the subsample to estimate the spread of the phenomena in the entire population. We analytically show that our estimates provide an upper bound and a lower bound on the spread of the disease in expectation. Finally, using a large high-resolution real-world mobility dataset, we experimentally show that our estimates are accurate in practice, while other methods that do not correctly account for co-locations between individuals result in entirely wrong observations (e..g,, premature herd-immunity) as the error is amplified over time.
\end{abstract}


\maketitle
\pagestyle{\vldbpagestyle}
\begingroup\small\noindent\raggedright\textbf{PVLDB Reference Format:}\\
\vldbauthors. \vldbtitle. PVLDB, \vldbvolume(\vldbissue): \vldbpages, \vldbyear.\\
\href{https://doi.org/\vldbdoi}{doi:\vldbdoi}
\endgroup
\begingroup
\renewcommand\thefootnote{}\footnote{\noindent
This work is licensed under the Creative Commons BY-NC-ND 4.0 International License. Visit \url{https://creativecommons.org/licenses/by-nc-nd/4.0/} to view a copy of this license. For any use beyond those covered by this license, obtain permission by emailing \href{mailto:info@vldb.org}{info@vldb.org}. Copyright is held by the owner/author(s). Publication rights licensed to the VLDB Endowment. \\
\raggedright Proceedings of the VLDB Endowment, Vol. \vldbvolume, No. \vldbissue\ %
ISSN 2150-8097. \\
\href{https://doi.org/\vldbdoi}{doi:\vldbdoi} \\
}\addtocounter{footnote}{-1}\endgroup

\ifdefempty{\vldbavailabilityurl}{}{
\vspace{.3cm}
\begingroup\small\noindent\raggedright\textbf{PVLDB Artifact Availability:}\\
The source code, data, and/or other artifacts have been made available at \url{\vldbavailabilityurl}.
\endgroup
}
\begin{sloppy}
\section{Introduction}
\textbf{Phenomena that Spread through Contact}. Viruses spread across a population through contacts and so do news, gossips, ideas and habits. Packages are passed between individuals when they meet to reach a destination, and pamphlets are given to individuals with the hope of reaching a wide audience. \textit{Physical contacts} are responsible for such phenomena passing from one individual to another. Their \textit{spread}, defined as how many people in a given population the phenomenon reaches, impacts our day to day lives, with COVID-19 as an on-going exhibit. Since the spread of such phenomena happens through contacts, it primarily depends on people's \textit{mobility patterns} (e.g., how people move in a city and how often they are co-located with others), as well as the characteristics of the phenomenon which determine how it can be passed on from one person to another, or the \textit{diffusion model} (e.g., the probability of transmission from one person to another given that they are co-located). The mobility pattern determines when and where contacts happen, while the diffusion model determines how a phenomenon spreads when there are contacts. 

\noindent\textbf{Role of Mobility Patterns}. Mobility patterns play a fundamental role in the spread of any phenomena in a population and its analysis. To shed more light on their role we consider two aspects of mobility. (1) \textit{Visits} form the \textit{location sequence} of an individual which contain the information about the locations an individual has been to. (2) \textit{Co-locations} between multiple individuals, a by-product of their visits, which contain the information on the contacts between individuals. The spread of a phenomena in a population depends on the second aspect, i.e., co-locations, since it determines which individuals were in contact and able to pass an item between them. This simple observation informs much of our later discussions.

\noindent\textbf{Benefits of Analyzing Spread}. The spread of the phenomena discussed has significance for both policy making and personal decision making. Given access to the mobility pattern of all individuals in a population and a diffusion model, we can simulate how a phenomena spreads in the population. This can be used to identify hotspots, compute location and individual risk-scores \cite{Rambhatla2020, Chiang2020, kiamari2020} and study various interventions or what-if scenarios \cite{kerr2020covasim, chang2020modelling, halloran2008modeling, ferguson2020report, ferguson2006strategies}. For instance, in the case of COVID-19, we can model the impact of wearing a mask as a change in the diffusion model (e.g., by lowering the probability of a transmission during a contact), and see how the spread differs from not wearing a mask, as done in \cite{eikenberry2020mask}. Having access to the mobility patterns for different cities at different times, allows us to perform these studies at a high resolution, e.g., for specific neighbourhoods and time periods.

\noindent\textbf{Mobility Patterns of a Population}. The benefits discussed above can only be materialized if we have access to mobility patterns for an entire population. However, that is not feasible in practice. Location sequences can be collected through cell-phones, but it is difficult, if not impossible, to convince every single individual to share their location with a single entity. Nevertheless, running simulations with millions of users and billions of locations is very computationally demanding. 

Meanwhile, location data of subsets of individuals, obtained through their cell-phones, has become available. For instance, Veraset \cite{veraset}, a data-as-a-service company, provides anonymized population movement data collected through GPS signals of cell-phones across the US. Such datasets provide high-resolution and detailed mobility patterns of parts of the population, but will likely never contain the entire population. It is important to be able to fully and correctly utilize such datasets to analyze the spread of a phenomenon over an entire population.

\noindent\textbf{Problem of Up-Sampling}. In this paper, we formalize and study the problem of estimating the spread of a phenomena in a population, given access to sub-samples of location visits of individuals in the population. Our goal is to estimate how many people in the whole population a phenomenon will reach, given a diffusion model and sample visits of the population. For instance, by observing only a portion of the population in a city, solving this problem allows us to estimate how much COVID-19 will spread under different intervention strategies (e.g., if people wear or do not wear masks) for the entire population.   

\noindent\textbf{Existing Solutions and Challenges}. A simple solution to the problem is to consider the spread in the sub-sampled population, and then scale the estimation to the whole population. However, we observed that such an approach vastly underestimates the spread because it ignores the co-locations between the unobserved individuals and the sampled individuals. In fact, the main challenge in solving the problem is accurately modeling co-locations between unobserved individuals and sampled individuals. We observed in our experiments that when this is not done accurately, the estimation can provide wrong infection patterns, e.g., estimating that the spread is stopping when the spread is actually increasing. This is because the underestimation gets amplified over time. For instance, underestimating the number of people who are currently infected in a population leads to further underestimating how many people the disease can be transmitted to in the future.
 
An alternative approach is to model the visits of the unobserved individuals using the visits of the sampled individuals, and use that model to infer co-locations between unobserved and sampled individuals. For instance, an approach can be to generate synthetic location trajectories, e.g., using \cite{ouyang2018non, feng2020learning}, based on the observed location sequences to create a larger dataset that contains both real and synthetic location sequences. However, we observed that this indirect formation of co-locations from synthetic location data has two limitations. First, the co-locations cannot be formed accurately because our model needs to generate extremely accurate synthetic locations, within a few meters (to generate correct co-locations), and for long periods of time (to be able to track infected individuals correctly). In addition, creating synthetic locations corresponding to real-world populations explodes the data size, rendering simulations at scale impractical. A detailed discussion of these methods is provided in Sec. \ref{sec:rel_work} and we experimentally evaluated a representative of such approaches in Sec. \ref{sec:exp} which confirmed the above observations. 

Effectively, due to lack of access to individual location data, agent based simulations that are used to assess various interventions for containing the spread of a disease \cite{kerr2020covasim, chang2020modelling, halloran2008modeling, ferguson2020report, ferguson2006strategies}, utilized across the world for the COVID-19 pandemic, use fixed contact matrices to generate co-locations between individuals. These contact matrices contain, at an aggregate level, the rate of contacts between different compartments in the population (e.g., the rate of contacts between people of age 10 with people of age 50 in a population), and are created through surveys and interviews in various parts of the world \cite{prem2020projecting, prem2017projecting, mossong2008social}. In the absence of individual location data, such contact matrices can be useful for studying the spread in a population at an aggregate level (e.g. for a country). However, since the contact matrices are static, i.e., do not change with time, and are created at an aggregate level, they do not take into account the spatiotemporal changes in mobility (e.g., change in mobility on a day-to-day basis, or for different neighbourhoods in a city). Thus, they cannot be used to provide accurate estimate of the spread at a particular point in time and space. We believe using real location data and an accurate method for estimating the spread can help empower the above-mentioned studies to better understand the impact of different interventions.  

\noindent\textbf{Our Approach}. We rigorously study the problem of up-sampling. Our approach is to statistically estimate the probability of a sampled individual getting infected. Such an estimate needs to take into account the probability of a sampled individual getting infected by unobserved individuals. Rather than modeling location visits of an individuals, from which co-locations can be indirectly inferred, we directly model co-locations between individuals. This follows our observation that the extra information associated with location visits (e.g., their exact Geo-coordinates and their entire sequence) is unnecessary for modeling co-locations, while modeling such extra information makes the model less accurate.

Our methods use a time-varying contact network \cite{shirani2012efficient} to model the co-location between individuals. The general approach is to use some statistics of the contact network, that can be estimated from sub-sampled individuals, to estimate the spread. We discuss two different ways this can be done. Our first approach, called Polling the Spreader, or PollSpreader, does this from the spreaders view: by modeling how many individuals a person can transmit the phenomena to. We observe that estimating statistics for this kind of modeling is difficult over long periods. Thus, we discuss our second approach, called Polling the Susceptible, or PollSusceptible, where we estimate the spread from a different perspective: we model different ways the phenomena can be transmitted to a particular individual. Using this approach, we provide lower and upper bounds on the spread of the phenomena in the whole population from the sub-sample. We experimentally observed that our estimates follow closely the spread in the whole population. Furthermore, our results show that the pattern of spread can be completely misjudged (e.g., showing early herd-immunity by mistake) if the co-locations are not estimated accurately.

\noindent\textbf{Contributions and Organization}. In this paper, we
\begin{itemize}
    \item Define the problem of estimating spread through sub-sampled visits (Sec. \ref{Sec:def});
    \item Present two novel methods, PollSpreader and PollSusceptible, that solve the problem accurately (Sec. \ref{sec:upsample}); 
    \item Theoretically study the problem and provide lower and upper bound estimates of the number of infected people in the population over time using PollSusceptible; and
    \item Experimentally show that our estimations are accurate in practice. (Sec. 4)
\end{itemize}

Furthermore, Secs. \ref{sec:rel_work} and \ref{sec:conc} discuss the related work and our conclusion, respectively.

\section{Problem Definition}\label{Sec:def}
We study the problem of the spread of a phenomenon across a population, where the phenomenon is passed from one person to another through their co-location. We assume only a sub-sample of the population can be observed, and our goal is to estimate the number of individuals in the true population who become subjected to the phenomenon. We next define the terminology used. We have summarized our terminology in Table \ref{tab:notation}.

\noindent\textbf{Mobility Pattern}. We consider a population consisting of $n$ individuals. For each individual, a visit provides the location of the individual at a particular point in time. Associated with each individual, $u$, is a \textit{visit sequence} or \textit{location sequence}, which is a sequence of the location of their visits at different times. For two consecutive visits, $c$ and $c'$ of an individual at times $t$ and $t'$, we assume the individual is at the location specified by $c$ from time $t$ to $t'$. As a result, given the visit sequence of $u$, we define $l_u$ to be a function returning the location of $u$ at every point in time. That is, for any $t$, $l_{u}(t)=(x_{t}^u,y_{t}^u)$ where $x_{t}^u$ and $y_{t}^u$ are the coordinates of the location of $u$ at time $t$. We abuse the notation and also refer to $l_u$ as $u$'s visit sequence. 

\begin{figure}[t]
    \centering
    \includegraphics[width=\columnwidth]{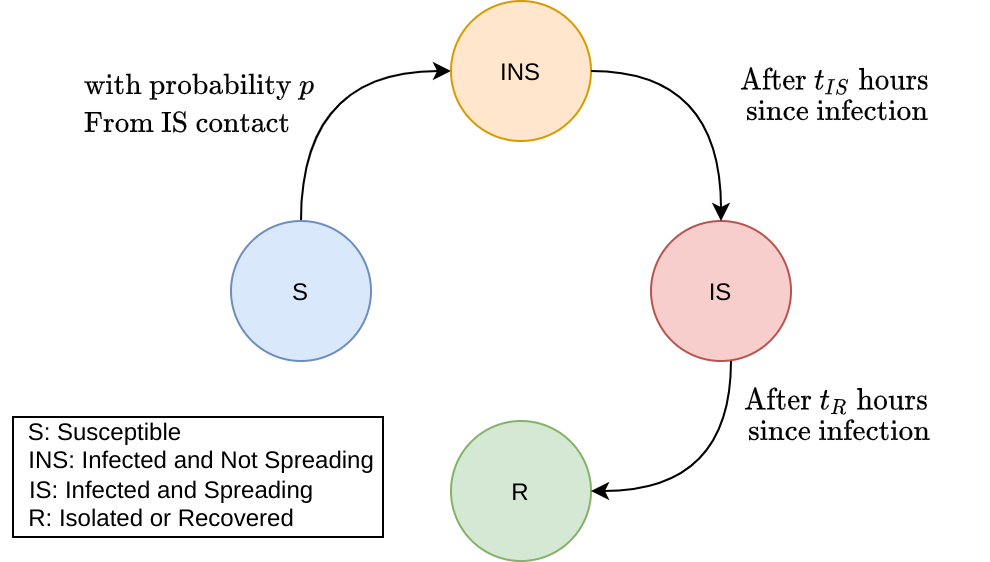}
    \caption{Diffusion Model}
    \label{fig:trans}
\end{figure}

\noindent\textbf{Diffusion Model}. Here we discuss how diffusion occurs over the population. For ease of discussion, and without loss of generality, we adopt a terminology commonly used when a disease spreads in a population and in particular the SIR model \cite{kermack1927} which is commonly used in epidemiology. The model is shown in Fig. \ref{fig:trans}. Each individual is in either the Susceptible (S), Infected (I) or Recovered (R) compartment. We refer to the compartment an individual belongs to as their \textit{status}. Recovered individuals are assumed to be immune to the disease (or deceased), and thus will not contract the disease anymore. Infected individuals are either able to spread the disease (called Infected and Spreading, or IS) or they are not (called Infected and not Spreading, or INS). Susceptible individuals can contract the disease from Infected and Spreading individuals. After an S individual, $u$, contracts the disease at time $t$, $u$ immediately become INS. Then, at time $t+t_{IS}$, where $t_{IS}$ is a positive number sampled from some distribution $\Theta_{IS}$, $u$ becomes IS. Furthermore, at time $t+t_{R}$, where $t_{R}$ is a positive number larger than $t_{IS}$ and sampled from some distribution $\Theta_{R}$, $u$ becomes R. 

Consider a Susceptible individual, $u$, and an Infected and Spreading individual, $v$, and their corresponding visit sequences $l_u$ and $l_v$. At any time, $t$, with probability $p=f(l_u, l_v, t)$, $u$ contracts the disease from $v$, where $f(l_u, l_v, t)$ is an application-dependent and user-defined function. The function $f(l_u, l_v, t)$ determines the probability of transmission of the disease from one person to another depending on the location of the individuals. Intuitively, there can be a non-zero probability of transmission when two individuals are close enough spatially for a long enough period of time.     

Finally, individuals are initially infected based on some probability distribution. Specifically, at time zero, a vector $c$ is sampled from a distribution $\Theta_{init}$, where $c\in\{0, 1\}^{n}$ and the $i$-th element of $c$ determines whether the $i$-th individual in the population is initially infected or not. If a user is initially infected, the user becomes an INS user and they transition to IS and R states following the same procedure discussed above. 

\noindent\textbf{Problem Statement}. Our goal is to estimate the extent of the spread over the population from just observing the visit sequences of a \textit{sub-sample} of the population. Specifically, consider a population $\mathcal{U}$ of $n$ individuals and a set $\mathcal{S}\subseteq \mathcal{U}$, where $\mathcal{S}$ is sampled from $\mathcal{U}$ uniformly at random. Each individual is sampled independently and with probability $p_s$. We use the term \textit{sub-sampled population} to refer to the set $\mathcal{S}$ and \textit{whole population} to refer to $\mathcal{U}$. We refer to individuals in $\mathcal{U}\setminus\mathcal{S}$ as unobserved individuals. 

\begin{definition}[Up-Sampling Infected Count Problem]
Give a sub-sampled population, $\mathcal{S}$, and the parameters of the diffusion model, for any given time $t$, the Up-Sampling Infected Count Problem is to return the expected number of individuals who have gotten infected in the whole population, $\mathcal{U}$, until time $t$.
\end{definition}

We use $EI_t$ to denote any estimation of the answer to the Up-Sampling Infected Count Problem. We can analogously define the problem for other compartments in the population (e.g., Susceptible or Recovered), but the general methodology towards solving the problem remains the same and thus we focus only on estimating the number of Infected individuals. 
\begin{table}[t!]
    \hspace*{-0.4cm}
    \begin{tabular}{c|c}
        Notation & Meaning  \\\hline
        $p_s$ & Probability of an individual begin sampled\\\hline
        $\theta_{IS}$ & Distribution of time it takes to become IS\\\hline
        $\theta_{R}$ & Distribution of time it takes to become R\\\hline
        $\theta_{init}$ & Distribution of initial infections\\\hline
        $\mu_{IS}$ & Time it takes to become IS for Sec. \ref{sec:upsample}\\\hline
        $\mu_{R}$ & Time it takes to become R for Sec. \ref{sec:upsample}\\\hline
        $d_{max}$ & Maximum distance for transmission for Sec. \ref{sec:upsample}\\\hline
        $t_{min}$ & Minimum duration for transmission for Sec. \ref{sec:upsample}\\\hline
        $p_{init}$ & Probability of an initial infection for Sec. \ref{sec:upsample}\\\hline
        $S_u$ & Indicator r. v. equal to one if $u$ is sampled\\\hline
        $X_u^t$ & Indicator r. v. equal to one if $u$ gets infected until time $t$\\\hline
        $\hat{p}_u^t$ & Estimate of $P(X_u^t=1)$ from sub-sampled population\\\hline
        $N(u, G_{s, t})$ & Neighbourhood of $u$ in $G_{s, t}$\\\hline
        $N_u$ & $N(u, G_{s, t})$, Neighbourhood of $u$ in $G_{s, t}$\\\hline
        $N_u^A$ & $N(u, G_{s, t})\setminus A$, Neighbourhood of $u$ in $G_{s, t}$ excluding A\\\hline
        $EI_t$ & An estimate for the Up-Sampling Infected Count Problem
    \end{tabular}
    \caption{Table of Notations}
    \label{tab:notation}
\end{table}

\section{Estimating the Spread}\label{sec:upsample}
Our goal is to estimate how many people in the whole population are infected at a point in time, $t$, by just observing the location of a sub-sample population. The problem would have been trivial had we had access to the true status of the sub-sampled population at time $t$. That is, at time $t$, if we knew exactly $k_t$ of the sub-sampled population are infected (or knew their exact probability of being infected), then an unbiased estimate of the number of infected individuals in the population would have been $\frac{k_t}{p_s}$. However, except for $t=0$, obtaining the correct value of $k_t$ is difficult, since a sampled individual may get infected by individuals who were not sampled (i.e., unobserved individuals). In this section, we first discuss a concrete diffusion model and introduce the necessary terminology  (Sec.~\ref{sec:term}) and illustrate the difficulty of obtaining such an estimate with two naive solutions (Secs.~\ref{sec:scale} and ~\ref{sec:frame_and_density}). We then present our methodology (Sec.~\ref{sec:graph_based}) and discuss how our solution can be applied to other diffusion models (Sec.~\ref{sec:generalizing}).

\subsection{Terminology and Diffusion Model}\label{sec:term}
\textbf{Diffusion Model}. For ease of discussion and concreteness, we present our methodology on a diffusion model with the following parameter setting. The parameters $\Theta_{IS}$ (time to spreading) and $\Theta_{IR}$ (time to recovery) are set to a distribution that returns $\mu_{IS}$ and $\mu_{R}$ respectively with probability 1, i.e. they are deterministic (e.g., in Fig.~\ref{fig:trans} $t_{IS}=\mu_{IS}$ and $t_{R}=\mu_{R}$). Furthermore, $\Theta_{init}$ (initial infections) is set such that every individual is initially infected independently with probability $p_{init}$. Finally, $f(l_u, l_v, t)$ is defined as follows. Intuitively, if an IS and an S individual are within $d_{max}$ of each other for at least $t_{min}$, then the IS individual will infect the S individual with probability $p_{inf}$, for user-defined parameters $p_{inf}$, $d_{max}$ and $t_{min}$. More formally, consider the time $t_1$ when the co-location between $u$ and $v$ starts. That is, $d(l_u(t_1), l_v(t_1))$ is at most the parameter $d_{max}$ and that right before $t_1$, $d(l_u(t), l_v(t))>d_{max}$. Let $t_2$ be the timestamp when the co-location between $u$ and $v$ ends. That is, $t_2$ is the first timestamp after $t_1$ such that $d(l_u(t_2), l_v(t_2))>d_{max}$. If the duration of the co-location, i.e., the time difference from $t_2$ to $t_1$, is at least a parameter $t_{min}$, then $f(l_u, l_v, t_{1})=p_{inf}$, for the parameter $p_{inf}$ denoting the probability of infection. $f(l_u, l_v, t)=0$ for all other timestamps. We use the term \textit{contact} to refer to co-locations within a distance $d_{max}$ that last for at least $t_{min}$ units of time. We discuss how our methodology applies to more general diffusion models in Sec.~\ref{sec:generalizing}. 

\noindent\textbf{Terminology}. For an individual $u$, the indicator random variable $S_u$ is equal to 1 if $u$ is sampled. Furthermore, the indicator random variable $X_u^t$ is equal to 1 if $u$ gets infected at a time less than or equal to $t$, and zero otherwise.

We refer to infections that were caused by transmission through $k$ individuals starting from an initial infections as \textit{$k$-hop infections}. That is, for a sequence of individual $<v_0, v_1, v_2, ..., v_k>$, we call an infection a $k$-hop infection if the individual $v_{k}$ was infected by $v_{k-1}$, $v_{k-1}$ was infected by $v_{k-2}$ and so on, and that $v_{0}$ was an initial infection (patient zero). That is, $v_0$ was infected at time $0$.

\vspace{-0.5cm}
\subsection{First Attempt: Scaling}\label{sec:scale} 
Our first simple solution is to use Monte Carlo simulation to estimate the spread of the disease in the sub-sampled population. 

\noindent\textbf{Methodology}. Given a sub-sample, the randomness in the spread is due to the randomness of the initial infections and the stochasticity of transmissions. To simulate the spread, a number of individuals are initially infected according to $\Theta_{init}$. Then, the visit sequences of the individuals are used to determine contacts between IS individuals and S individuals, and for every such contact the S individuals gets infected with probability $p_{inf}$. In such a simulation, at time $t$, define $k_t$ to be the number of people that are infected. We run the simulation $r$ times, obtaining the estimate $k_t^i$ for time $t$ from the $i$-th run. Then we return $\frac{1}{p_s}\frac{\sum_ik_t^i}{r}$ as our estimate of expected number of infections at time $t$. Here $\frac{1}{r}$ is for taking the mean of the $r$ simulations, and $\frac{1}{p_s}$ is for scaling up from the sub-sampled population to the full population.

\noindent\textbf{Analysis}. Such an estimate provides a biased estimate of the expected number of infections that occur after the initial infections. It specifically underestimates the number of infections for any time $t>0$. To analyze the accuracy, for a specific run of the simulation, let $\hat{X}_u^t$ be a random variable denoting if an individual $u$ gets infected by other sampled individuals until time $t$. Then, the estimate, $EI_t$, at time $t$ is 

\begin{align}\label{eq:scale}
    EI_t=\frac{1}{p_s}\sum_u S_u\hat{X}_u^t
\end{align}

First, note that at time $t=0$, the estimate is unbiased. That is, taking the expected value of Eq.~\ref{eq:scale} for $t=0$, we obtain $E[S_u\hat{X}_u^0]=E[S_u]E[\hat{X}_u^0]=p_s\times p_{init}$ and the expected number of infections in the whole population is $n\times p_{init}$. Thus, the estimate $\frac{1}{p_s}\sum_u S_u\hat{X}_u^0$ is an unbiased estimate of the number of infections at time $t=0$. 

However, consider $\hat{X}_u^t$ for $t>0$. Observe that $P(\hat{X}_u^t=1)$ can be less than $P(X_u^t=1)$. This is because, with some probability, individuals with whom $u$ has contacts may not be sampled, reducing the probability of $u$ getting infected from the sampled individuals. In terms of the Monte Carlo simulation, this can be seen as the case of \textit{false negatives} in the simulation. That is, even assuming initial infections and the transmission are deterministic, $u$ may get infected from someone in the whole population, but may not get infected from anyone in the sub-sample population, because the transmission of the disease to $u$ in the whole population is through unobserved individuals. 

\noindent\textbf{Observations}. Our experiments show that this underestimation is significant in practice, amplified because it cascades through multi-hop infections (e.g., second-hop infections are underestimated, which in turn makes the estimate of third-hop infections worse). Simply scaling up to the full population does not work, which we show in the experiments in Sec.~\ref{sec:exp}. This observation suggests that we need to account for the visit sequence that were \textit{lost} when sub-sampling.

\begin{figure}
    \centering
    \includegraphics[width=0.8\columnwidth]{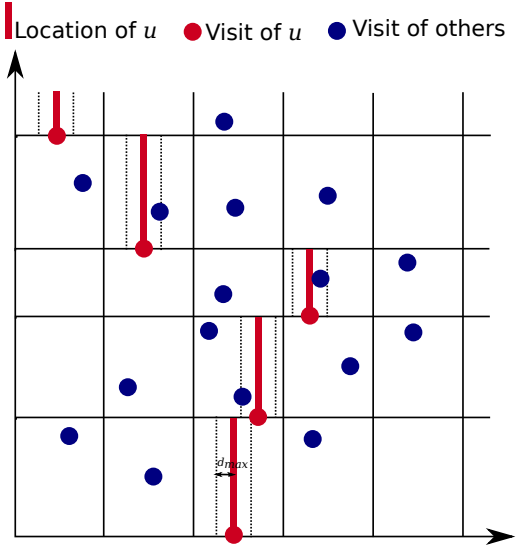}
    \caption{Modeling distribution of visits for an individual $u$. The horizonal axis is longitude, and the vertical axis is time. For ease of display, we do not show latitude. A red vertical line represents a visit over time by an individual u, and a blue dot represents the presence of another individual}
    \label{fig:histogram}
\end{figure}

\subsection{Second Attempt: Location Modeling}\label{sec:frame_and_density}
\subsubsection{Framework\nopunct}\label{sec:framework} \hfill\\Sec.~\ref{sec:scale} illustrates the need for accounting for the unobserved individuals. Our general framework is to try to estimate $P(X_u^t=1)$. We denote by $\hat{p}_u^t$ an estimate of $P(X_u^t=1)$. To solve the issues arising in Sec.~\ref{sec:scale}, we need to account for the unobserved individuals when calculating $\hat{p}_u^t$. This algorithm differs from our first attempt in how we model the unobserved individuals to calculate $\hat{p}_u^t$. After we obtain $\hat{p}_u^t$ for all sub-sampled individuals, our final estimate is 
\begin{equation}\label{eq:est_density}
    EI_t=\frac{1}{p_s}\sum_u S_u\hat{p}_u^t
\end{equation}

\noindent Comparing Eq.~\ref{eq:est_density} with Eq.~\ref{eq:scale}, the main difference in this approach is the flexibility it allows in estimating $P(X_u^t=1)$.

\subsubsection{Second Attempt: Density-Based Estimation\nopunct}\label{sec:density}\hfill\\
Our second attempt models the distribution of the visits of the unobserved individuals. It uses the model to calculate the probability of a sub-sampled individual being infected by unobserved individuals. This is subsequently used in calculating $\hat{p}_u^t$. 

\noindent\textbf{Methodology}. Our approach follows four steps: (1) Modeling the probability distribution of the visits of the unobserved population, (2) calculating the probability of $u$ getting infected by a single visit of an unobserved individual when $u$ is in a particular cell, (3) estimating probability of $u$ getting infected when it's in a particular cell, and  (4) estimating $P(X_u^t=1)$ by taking into account past visits of $u$.

\textit{Step (1), Modeling Visit Distribution}: We discretize time and space using a grid. We use a uniform grid across space, but a per user grid across time. That is, for each user, we model the visit distribution with a different histogram, but all the histograms model the same distribution, i.e., the distribution of the check-ins of all the individuals. The grid across time uses event-based discretization for each individual. In particular, for an individual $u$, let $C=<t_0, t_1, ...>$ be the sequence of times we have observed a visit. According to the discussion in Sec.~\ref{Sec:def}, $u$ is in the same location during time $t_i$ to $t_{i+1}$ for all $i$. Thus, for user $u$, we discretize time based on $C$. The choice of griding for time simplifies our analysis as every visit now corresponds to a cell, but other griding strategies can be similarly applied. For each grid cell $(t, i, j)$, let $n_{t, i, j}$ be the number of observed visits that fall into that cell, where we also count the individuals who are already in the cell $g=(t, i, j)$ when $u$ enters the cell, to avoid the scenarios where no new check-ins are observed but infections happened through individuals already in the cell (that is, if the beginning time of the cell is $a$ and the observed individual $v$ is already in the cell $g$ at time $a$, we increment $n_{t, i, j}$ by one). Fig. \ref{fig:histogram} shows the grid used for modeling for an individual $u$, where $n_{t, i, j}$ is calculated by counting the number of blue circles. Note that for different individuals, we will have different horizontal partitioning in the figure. We define 

\begin{equation}\label{eq:p_loc}
    p_{t, i, j}=\frac{n_{t, i, j}}{\sum_{i', j'}n_{t, i', j'}}
\end{equation}

\noindent to model the probability of a visit falling into the cell $(t, i, j)$. Furthermore, we assume visits inside a cell are distributed uniformly in space. In this step, we do not model the amount of time that an observed individual $v$ (blue dot in Figure~\ref{fig:histogram}) is in the cell.

\textit{Step (2), Prob. of Infection from a Single Visit}: Consider a visit $c$, corresponding to a grid cell $g=(t, i, j)$ of a sampled susceptible individual, $u$. Assume we know that a particular visit $c'$ by some unobserved infected individual happens when $u$ is in $g$. We need to calculate the probability of $c'$ being within $d_{max}$ of $c$ and lasting for at least $t_{min}$. Let $p_{c'}$ be the probability that $c'$ causes $u$ to get infected. Moreover, let $p_{long\_enough}$ be the probability of a co-location lasting at least $t_{min}$.

\begin{align}
    p_c=p_{t, i, j}\times \frac{\pi d_{max}^2}{\text{cell area}}\times p_{long\_enough}
\end{align}

\noindent where $\frac{\pi d_{max}^2}{\text{cell area}}$ is the probability of co-location given that the visits are in the same cell (ignoring the edge cases where co-locations happen from two different cells). In Fig. \ref{fig:histogram}, $p_{t, i, j}\times \frac{\pi d_{max}^2}{\text{cell area}}$ is analogous to calculating the probability of a new blue circle (i.e., visit by others) falling within the dashed line of the red rectangle (i.e., within $d_{max}$ of location of $u$) for a particular time. We model the duration an individual stays at a location with an exponential random variable, so $p_{long\_enough}$ is the probability of the random variable being more than $t_{min}$. 

\textit{Step (3), Prob. of Infection in a Cell:} If there are $N_c$ visits by unobserved infected users for the duration that $u$ is in the cell, probability of the user getting infected during that time period is 
\begin{equation}
p_t=1-(1-p_c)^{N_c}    
\end{equation}

To estimate $N_c$, we estimate (a) the number of unobserved IS individuals, $N_{IS}$ and (b) the average number of visits per IS individual $N_{c\text{ per IS}}$. Then we estimate $N_c=N_{IS}\times N_{c\text{ per IS}}$. For (a), assume that the current time is $t$. An IS individual must have been infected some time between $t-\mu_{IS}$ and $t-\mu_{R}$. We use our estimate of the number of infected individuals for times $t-\mu_{IS}$ and $t-\mu_{R}$ to calculate the number of current IS individuals. For (b), we use $N_{c\text{ per IS}}=\frac{\sum_{i, j}n_{t, i, j}}{p_s}$, where $\sum_{i, j}n_{t, i, j}$ is the total number of observed visits during the time $u$ is in cell $g$, and we scale it by $\frac{1}{p_s}$ to get to the whole population (here we have assumed that the average number of visits per IS individuals is the same as the average number of visits for all individuals).

\textit{Step (4), Prob. of Infection until current time}: Finally, if $u$ enters $N_t$ grid cells until time $t$, the probability of $u$ getting infected until time $t$ is estimated as

\begin{equation}
    \hat{p}_u^t=1-\prod_{i=0}^{N_t}(1-p_i)
\end{equation}

\noindent\textbf{Observations}. We experimentally observed that this approach still leads to underestimation (see Sec.~\ref{sec:exp}). This happens for two reasons. First, on one hand, there are inevitable inaccuracies in modeling the visits spatially due to modeling assumptions. For instance, the use of a grid assumes locations that are spatially close have similar visit density. This may not be true in practice, because for instance a church may exist in a residential area. Although this may be possible to address by increasing the granularity of the grid cells, doing so will require a large set of sampled users to avoid over-fitting (as otherwise a visit will have zero probability of falling into most gird cells). On the other hand, to accurately model co-locations, a model that is accurate to within a few meters is needed. Second, the temporal correlations between the visits needs to be modeled. For instance assume that an infection happens in a particular grid-cell, $g$ at time $t$. It means that an unobserved IS user, $v$, was in $g$ at time $t$. This changes the probability distribution of the locations where $v$ can be at time $t+1$. Overall, both points imply that an approach that aims to up-sample the infection data by modeling the location sequences requires a very precise modeling of the location sequences both spatially and temporally. We present accuracy results from this model in Sec.~\ref{sec:exp}.

An interesting observation is that our end-goal is to use the model of the location sequences to find possible co-locations between individuals. That is, the spatial information associated with a location sequence is only a means to an end, but is not necessary. Instead, we can directly model the possible co-locations.

\subsection{Third attempt: Co-location Modeling}\label{sec:graph_based}
Motivated by the above observations, we aim at directly modeling the contacts between the individuals. We present two approaches that provide bounds on the expected spread of the disease in the population. Our first approach, PollSpreader, models the problem from the spreaders' view, i.e., aims at calculating how many people will get infected given a number of spreaders (Sec.~\ref{sec:upsampling:spreader}). We observed that, although this approach works well when modeling first-hop infections, up-sampling for multi-hops becomes difficult. Therefore, we present our second, PollSusceptible, approach which looks at the problem from the susceptible  view, e.g., directly calculates the probability of a susceptible individual getting infected (Sec.~\ref{sec:upsampling:susceptible}). Both approaches use a time-dependent contact network which we first describe in Sec.~\ref{sec:contact_network}. 

\begin{figure}
    \centering
    \includegraphics[width=\columnwidth]{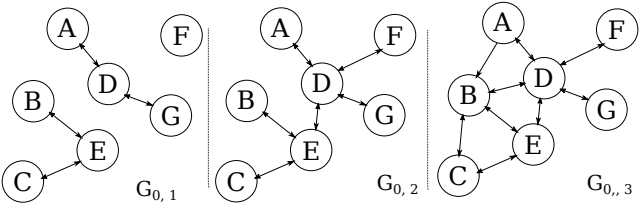}
    \caption{A contact network}
    \label{fig:contact_network}
\end{figure}
\begin{figure}
    \centering
    \includegraphics[width=\columnwidth]{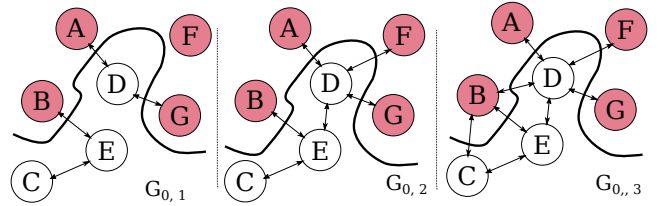}
    \caption{A cut on the network (Red: Initial infections, White: Rest of the individuals)}
    \label{fig:network_cut}
\end{figure}

\subsubsection{Dynamic Contact Network\nopunct}\label{sec:contact_network}\hfill\\
We use a dynamic weighted directed graph to model contacts between individuals. Specifically, for the contacts from time $s$ until time $t$, we build a graph $G_{s,t}=(V, E_{s, t})$, where $V$ contains a node for every individual and $E_{s, t}$ contain edges between users that are in contact during the time period [s, t]. Specifically, there is an edge from node $u$ to node $v$ if there exists a co-location between them. Furthermore, the weight on the edge $(u, v)$, denoted by $w_{u, v}$, is the probability that $v$ gets infected from $u$ given that $u$ is infected at time $s$. Fig.~\ref{fig:contact_network} shows an example of a contact network. We use $\hat{G}_{s, t}=(\hat{V}, \hat{E_{s, t}})$ to refer to a contact network that is built from a sub-sample of the population. $\hat{G}_{s, t}$ will have fewer nodes than $G_{s, t}$.

\subsubsection{PollSpreader: Up-Sampling using Spreaders' View\nopunct}\label{sec:upsampling:spreader}\hfill\\
We study how many first-hop infections can occur based on an initial number of infections. Our approach follows the framework of Sec.~\ref{sec:framework}, and we aim at calculating $\hat{p}_u^t$. We first explain how the contact network can be used to do this if we have access to the entire population. Then, we discuss how the approach can be adjusted when we only observe a sub-sample of the population. Finally, we discuss the difficulty in extending the approach to multi-hop infections. We refer to this method as Polling the Spreader or PollSpreader.

\noindent\textbf{Calculating $\hat{p}_u^t$ from the Whole Population }. Consider the case where we have access to the whole population. Our approach works as follows. (1) We calculate the total number of contacts between initial infections and the susceptible population. Then, (2) we use that to estimate the probability of a susceptible individual getting infected. 

\textit{Step (1), Calculate Number of Infectious Contacts:} Here, we estimate the number of \textit{infection events}, informally defined as when an initially infected user, $v$, has contact with an individual $v$ who was not initially infected. More formally, define the infection random variable $I_{u, v}^t$ as an indicator random variable which is equal to one if $u$ is an initially infected person and $u$ infects $v$ until time $t$ given that no other individual infects $v$. We call it an \textit{infection event} when $I_{u, v}^t=1$. Our goal is to estimate how many \textit{infection events} occur in total in the whole population by calculating $E[\sum_{u, v}I_{u, v}^t]$.

Calculating $E[\sum_{u, v}I_{u, v}^t]$ can be formulated as estimating the expected total weight of the edges crossing a random graph cut on the graph $G_{0, t}$. Fig. \ref{fig:network_cut} shows how this is done for a specific realization of initial infections, where Fig. \ref{fig:network_cut} is obtained by a random selection of initial infections from Fig. \ref{fig:contact_network}, that is, the red nodes are randomly selected initial infections. Any such random selection defines a cut on the graph for every timestamp. The goal is to estimate the expected number of edges crossing such a cut. Specifically, consider the following random cut, where a node is in the set $I$ if it is initially infected, and in the set $S$ if it's not. Every cut corresponds to an initialization of the infections. Given a cut, the expected number of infection events until time $t$ is equal to the total weight of the edges crossing the cut from $I$ to $S$ (for each different $t$, we consider the cut on a different graph, as $G_{0, t}$ changes with time). Thus, the expected number of infection events is equal to the expected total weight of the edges crossing the cut from $I$ to $S$. For every edge $(u, v)$, assign the random variable $Z_{u, v}$ to be equal to $w_{u, v}$ if $u\in I$ and $v\in S$, and zero otherwise. We have that 

\begin{equation}\label{eq:spreader:cut}
    E[Z_{u, v}]=p_{init}\times (1-p_{init})\times w_{u, v}
\end{equation}

and the expected total weight of the edges crossing the cut is 

\begin{equation}
 E[\sum_{u, v}I_{u, v}^t]=e_c^t=\sum_{(u, v)\in E_{0, t}}p_{init}\times (1-p_{init})\times w_{u, v}   
\end{equation}

\textit{Step (2), Estimating $\hat{p}_u^t$ from Infection Events:} We assume the infection events occur uniformly at random across the susceptible population, or in other words, the edges from $I$ fall uniformly at random over the nodes in $S$. Although more sophisticated modelling approaches may be possible, we observed that this assumption works well in practice. Thus, given $Z_{u,v}$ for all $u$ and $v$, the probability of a node in $S$ having at least one edge incident on them is $1-(1-\frac{1}{|S|})^{\sum Z_{u, v}}$. We estimate this quantity by $1-(1-\frac{1}{E[|S|]})^{e_c^t}$. Finally, an individual either gets infected initially (with probability $p_{init}$), or does not get infected initially but gets infected through its contact with $u$. Hence, we estimate probability of getting infected as 

\begin{equation}\label{eq:spreader_est}
    \hat{p}_u^t= p_{init}+(1-p_{init})(1-(1-\frac{1}{E[|S|]})^{e_c^t})
\end{equation}

\noindent\textbf{Calculating $\hat{p}_u^t$ from the Sub-Sampled Population }. An interesting advantage of this modelling is how easily it can be adjusted when we only have access to a sub-sample of the population. In Eq.~\ref{eq:spreader_est}, we only need to be able to estimate $E[e_c]$. To do so we need to slightly modify our graph cut formulation. Specifically, let $\hat{I}$ be the set of nodes that are sampled and initially infected, and let $\hat{S}$ be the set of nodes that are sampled and not initially infected. Furthermore, let $\hat{Z}_{u, v}$ be equal to $w_{u, v}$ if $u\in \hat{I}$ and $v\in \hat{S}$, and zero otherwise. We have that 

\begin{equation}\label{eq:cut_est}
E[\hat{Z}_{u, v}]=p_s^2\times p_{init}\times (1-p_{init})\times p_{u, v}    
\end{equation}

Comparing Eq.~\ref{eq:cut_est} with Eq.~\ref{eq:spreader:cut}, we observe that adjusting for sampling can now be done just by scaling. Thus, an unbiased estimate of the total expected number of edges crossing the cut is $\frac{\sum_{u, v}\hat{Z}_{u, v}}{p_s^2}$. So

\begin{equation}
 E[\sum_{u, v}I_{u, v}^t]=\hat{e}_c^t=\frac{1}{p_s^2}\sum_{(u, v)\in \hat{E}_{0, t}}p_{init}\times (1-p_{init})\times w_{u, v}   
\end{equation}

Putting everything together, our final estimate replaces $e_c^t$ in Eq.~\ref{eq:spreader_est} with $\hat{e}_c^t$.

\noindent\textbf{Challenge for Multi-Hop Infections}. The main challenge associated with this method is generalizing it to more than one-hop neighbours. The contact network can be generalized to a \textit{reachability network} \cite{shirani2012efficient}, such that there is an edge from a node $u$ to another node $v$ if $u$ can infect $v$ through multi-hop infections (and not only direct contacts). However, the main challenge is taking the sampling into account, because the probability of an edge being sampled now depends on which of the intermediary nodes are sampled, and how many different $k$-hop paths from $u$ to $v$ exist in the population. Thus, we need a method that accounts for the paths in a multi-hop infection between initial infections and unobserved secondary infections.

\begin{algorithm}[t]
\begin{algorithmic}[1]
\REQUIRE Sequence of $k+1$ individuals, $s$, $s=<v_0, v_1, ..., v_{k}>$ and infection time of $v_{k}$, $t_{I}$
\ENSURE $P(T_s^t=1|v_k\text{ gets infected at }t_i)$
\IF {$k==0$}
    \RETURN 1
\ENDIF
\STATE $t_{IS}\leftarrow t_I+\mu_{IS}$\label{alg:trans:t_is}
\STATE $t_{R}\leftarrow t_I+\mu_{R}$\label{alg:trans:t_r}
\STATE $c_{v_{k}, v_{k-1}}\leftarrow$ list of contacts between $v_k$ and $v_{k-1}$ from $t_{IS}$ to $t_{R}$ ordered by time\label{alg:trans:contact}
\STATE $p\leftarrow 0$
\FOR {$c_i$ in $c_{v_{k}, v_{k-1}}$}
    \STATE $t_c \leftarrow$ time $c_i$ occurs
    \STATE $p_{inf\_by\_c}\leftarrow calc\_prob\_transmision(<v_0, ..., v_{k-1}>, t_c)$
    \STATE $p\leftarrow p+p_{inf\_by\_c}\times p_{inf}\times (1-p_{inf})^{i-1}$
\ENDFOR
\RETURN $p$
\end{algorithmic}
\caption{Calculating Prob. of Transmission Through a Path, $calc\_prob\_transmision(s, t)$}\label{alg:transmission_prob}
\end{algorithm}

\begin{algorithm}[t]
\begin{algorithmic}[1]
\REQUIRE Sequence of $k+1$ individuals, $s$, $s=<v_0, v_1, ..., v_{k}>$ and current time $t$ 
\ENSURE Lower and upper bounds estimates of $P(T_s^t=1)$
\STATE $p_{not\_l}\leftarrow 1$
\STATE $p_{not\_u}\leftarrow 1$
\FOR {$v'$ in $N(v_k, \hat{G}_{0, t})\setminus\{v_0, ..., v_k\}$}
    \STATE $p_{v'\_l}, p_{v'\_u}\leftarrow calc\_prob\_inf(<v_0, ..., v_{k}, v'>)$\label{alg:inf:rec}
    \STATE $p_{not\_l}\leftarrow p_{not\_l}\times (1-p_{v'\_u})$ 
    \STATE $p_{not\_u}\leftarrow p_{not\_u}\times (1-p_{v'\_l})$ 
\ENDFOR
\STATE $p_{starts\_at\_v_k}\leftarrow calc\_prob\_transmision(s, 0)$
\STATE $\hat{p}_{l}\leftarrow p_{starts\_at\_v_k}+(1-p_{starts\_at\_v_k})\times(1-(p_{not\_u})^{c_u})$
\STATE $\hat{p}_{u}\leftarrow p_{starts\_at\_v_k}+(1-p_{starts\_at\_v_k})\times(1-(p_{not\_l})^{c_l})$
\RETURN $\hat{p}_{l}, \hat{p}_{u}$
\end{algorithmic}
\caption{Calculating Prob. of Infection by a Path, $calc\_prob\_inf(s, t_{I})$}\label{alg:inf_prob}
\end{algorithm}

\subsubsection{PollSusceptible: Up-Sampling using Susceptibles' View\nopunct    }\label{sec:upsampling:susceptible}\hfill\\ To account for unobserved secondary infections, here, we aim at directly modeling the probability of a susceptible person getting infected. This is our second graph-based approach, which proved to be the most accurate in our experiments. By making some independence assumptions, this allows us to naturally model multi-hop infections. We refer to this method as Polling the Susceptible or PollSusceptible.

\noindent\textbf{Calculating $\hat{p}_u^t$ from the Whole Population}. We again start by assuming we have access to the entire population and use the same contact network discussed before. The general idea is to observe that an individual, $u$, who got infected was either initially infected, or got infected through some of its contacts that are modeled by the contact network. If $u$ got infected through a contact, $v$, the same logic recursively applies. That is, $v$ was either initially infected or got infected through a contact. To state this formally, we first introduce more terminology, then discuss the case of getting infected through neighbours and finally detail the recursion step.

\textit{Neighbourhood Terminology and Notation}. Recall that the contact network represents, for each individual, $u$, if they are infected, whom they can infect and with what probability. Consider $v\in N(u, G_{s, t})$, where $N(u, G_{s, t})$ is the set of neighbours of $u$, that is the set  of nodes, $v$, in $G_{s, t}$ such that there is an edge from $u$ to $v$ in $G_{s, t}$. $N(u, G_{s, t})$ represents the set of possible first-hop infections caused by $u$, given that $u$ starts spreading the disease at time $s$ and recovers at time $t$. We use the shorthand notation $N_{u}$ to refer to $N(u, G_{s, t})$ when the contact network in question is clear from the context, and we use the notation $N_u^A$, for some set $A$, to denote $N(u, G_{s, t})\setminus A$.

\textit{Calculating $\hat{p}_{v_0}^t$ through first-hop neighbours}. Our goal is as before, calculating $\hat{p}_{v_0}^t$ for all users $v_0$ following the framework of Sec.~\ref{sec:framework}. To do so, for individuals $v_0$ and $v_1$, let $T_{v_0, v_1}^t$ be the indicator random variable equal to one if the disease is transmitted from $v_1$ to $v_0$ until time $t$. 
Now, the event that $v_0$ does not get infected until time $t$, $X_{v_0}^t=0$, can be decomposed as the intersection of two events: one that $v_0$ does not get infected initially, $X_{v_0}^t=0$, and that $v_0$ does not get infected by any of its contacts until time $t$, written as $T_{v_0, v_1}^t=0$ for all $v_1\in N_{v_0}$. The idea can be illustrated using Fig.~\ref{fig:contact_network}. For instance, for $E$ to get infected until time $3$, it either has to be initially infected, or not initially infected but get infected from $B$, $C$ or $D$

Using this formulation, the probability that $v_0$ gets infected until time $t$ is 
\begin{equation}\label{eq:first_hop}
    \hat{p}_{v_0}^t=p_{init}+(1-p_{init})(1-\underbrace{P(T_{v_0,v_1}^t=0,\forall v_1\in N_{v_0})}_{p_{v_0}^N}
\end{equation} 

\noindent where the two events of $v_0$ either gets initially infected (first term), or it does not get initially infected but gets infected through transmission (second term) are accounted for. Denote $p_{v_0}^N=P(T_{v_0,v_1}^t=0,\forall v_1\in N_{v_0})$ for ease of reference. Here, we make the modeling assumption that the random variables $X_{v_0, v_1}^t$ are independent for all $v\in N_{v_0}$. Using the assumption, 

\begin{align}\label{eq:indep}
p_{v_0}^N=\prod\limits_{v_1\in N_{v_0}}P(T_{v_0, v_1}^t=0)
\end{align}

\noindent we acknowledge that this may not be necessary true for multi-hop infections, but we observed that the introduced error is negligible in practice per our experiments (see Sec.~\ref{sec:exp}). 

Combining Eqs.~\ref{eq:first_hop} and \ref{eq:indep}, the problem of calculating $\hat{p}_{v_0}^t$ is now reduced to calculating $P(T_{v_0, v_1}^t=0)$ for each $v_1$.

\textit{Recursion for Multiple Hops}. To calculate $P(T_{v_0, v_1}^t=0)$, a similar logic can be recursively applied. That is, $T_{v_0, v_1}^t=0$ can occur only if $v_1$ is initially infected, or if it is infected by another one of its neighbours. Continuing with our example, in Fig.~\ref{fig:contact_network}, for $E$ to be infected by $D$ until time 3, $D$ must have either been infected initially or gotten infected from $A$, $F$ or $G$. Also note that for $E$ to be infected by $B$ until time 2, $B$ must have gotten infected initially. 

Specifically, let $T_{v_0, v_1, v_2}^t$ be the random variable equal to one if $v_0$ gets infected until time $t$ by $v_1$ who gets infected by $v_2$, $v_2 \in N_{v_1}^{\{v_0\}}$. Thus, we can write 

\begin{align}
    P(T_{v_0, v_1}^t=1)&=p_{init}P(T_{v_0, v_1}^t=1|X_{v_1}^0=1)+\nonumber\\&(1-p_{init})(1-\prod_{v_2 \in N_{v_1}^{\{v_0\}}}P(T_{v_0, v_1, v_2}^t=0))
\end{align}

More generally, $P(T_s^t=1)$, when $s=<v_0, ..., v_k>$ can always be recursively calculated as

\begin{align}\label{eq:calc_inf_path}
    P(T_s^t=1)&=p_{init}\underbrace{P(T_{s}^t=1|X_{v_k}^0=1)}_{\text{full-path transmission}}+\nonumber\\&(1-p_{init})(1-\prod_{v_{k+1} \in N_{v_k}^ s}\underbrace{P(T_{s,v_{k+1}}^t=0}_{\text{recursion}}))
\end{align}

\noindent where $s,v_{k+1}$ denotes the concatenation of $v_{k+1}$ to $s$. A simple base-case for the recursion is when all the neighbours of $v_k$ are already on the path from $v_0$ to $v_k$ (in practice when estimating the spread until time $t$, we are generally able to stop the recursion based on the number of hops. This is because we know the time it takes for a given disease to be transmitted during each hop, and thus the number of hops that can happen until time $t$ can be estimated). 

The term labeled \textit{recursion} in Eq.~\ref{eq:calc_inf_path} shows the recursive calculation. The term labeled \textit{full-path transmission} is the probability of transmission happening given a complete path, e.g., the path starts with a patient zero, $v_k$, who was initially infected and continues until $v_0$. The probability of a full-path transmission can be calculated by considering the contacts between every pair of individuals, $v_i$ and $v_{i-1}$, on the path as explained below. Before moving on to the details of this calculation, we note that using Eqs.~\ref{eq:calc_inf_path} and~\ref{eq:indep}, we can exactly calculate Eq.~\ref{eq:first_hop}, which provides our estimate of $p_{v_0}^t$.

\textit{Calculating Probability of Full-path Transmissions}. Algorithm~\ref{alg:transmission_prob} shows how this is done. The general idea is to divide the event $T_{s}^t=1$, for $s=<v_0, ..., v_k>$, into a set of mutually exclusive events and sum up the probabilities of each event to calculate the final probability. Particularly, let $c_{u, v}$ be the list of contacts between $u$ and $v$ ordered by time, let $W(u, v, c, t_i)$, for $c\in c_{u, v}$ denote the event that $u$ infects $v$ through contact $c$ given that $u$ was infected at time $t_i$. Observe that the event $E=T_{s}^t=1|X_{v_k}^0=1$ is the same as $E=\cup_{c_i\in c_{v_{k}, v_{k-1}}}(T_{s\setminus v_k}^t=1\land W(v_k, v_{k-1}, c_i, 0))$ and that each event is mutually exclusive (because an individual can get infected at most once), where $s\setminus v_k$ is the sequence of elements in $s$ excluding $v_k$.

Note that 
\begin{align*}
  &P(T_s^t=1\land W(v_k, v_{k-1}, c_i, 0))=\\&\underbrace{P(T_{s\setminus v_k}^t=1|W(v_k, v_{k-1}, c, 0))}_{\text{recursion}}\times \underbrace{P(W(v_k, v_{k-1}, c_i, 0)}_{(1-p_{inf})^{i-1}\times p_{inf}}) 
\end{align*}

The first term is calculated recursively. To calculate the second term, recall that $c_i$ is the $i$-th contact between $v_k$ and $v_{k-1}$ since $v_{k-1}$ became IS. Thus, for $v_{k-1}$ to become infected through $c_i$, it has to be true that none of the previous $i-1$ contacts caused infection, and that $c_i$ did. Since each of the events are independent, the probability of this happening can be calculated as $(1-p_{inf})^{i-1}\times p_{inf}$.

\noindent\textbf{Calculating $\hat{p}_u^t$ from the Sub-Sample Population}. Now assume we only have access to a sub-set of the population. Calculating full-path transmission probabilities can be done in the same way as before (Alg.~\ref{alg:transmission_prob}). However, not all the neighbours of a user $v_0$ are sampled. That is, in Eq.~\ref{eq:first_hop}, we do not have access to $N(v_0, G_{0, t})$, but only its sub-sample $N(v_0, \hat{G}_{0, t})$. Thus, we need to adjust Eq.~\ref{eq:indep} to estimate $p_{v_0}^N$ from our sub-samples. We denote by $\hat{p}_{v_0}^N=P(T_{v_0,v_1}^t=0,\forall v_1\in N(v_0, \hat{G}_{0, t}))$, i.e., $\hat{p}_{v_0}^N$ this is our estimate of $p_{v_0}^N$ from the sub-sampled population.
The challenge in making such an estimate is that we are estimating the product across a population using a sub-sample. Providing unbiased estimates of summations using sub-samples can be easily achieved by merely scaling the sample statistic, but such an approach does not work for estimating products. This is because there is no simple relationship between $E[\hat{p}_{v_0}^N]$ and $p_{v_0}^N$. Instead, to be able to compute reliable estimates, we provide lower and upper bounds on $p_{v_0}^N$ as discussed in the following theorem. The theorem makes use of the quantity $p_{min}$, which is the largest number such that $p_{min}\leq P({T}_{u, v}^t=0)$, and is a bound on the probability of transmission.

\begin{theorem}\label{thm:subsample}
Let $c_l=\frac{1}{ps}$. We have that
$$E[(\prod_{v_1\in N(v_0, G_{0, t})}P(T_{v_0, v_1}^t=0)^{S_{v_1}})^{c_l}] \leq p_{v_0}^N$$
Furthermore, let $c_u$ be the solution to $c_up_s+\frac{c_u^2\log(p_{min})}{8}=1$. If $p_s\geq\sqrt{\frac{-log(p_{min})}{2}}$, we have that 
$$p_{v_0}^N\leq E[(\prod_{v_1\in N(v_0, G_{0, t})}P(T_{v_0, v_1}^t=0)^{S_{v_1}})^{c_u}]$$
\end{theorem}

\textit{Proof Sketch}. We take the log of $p_{v_0}^N=\prod_{v_1\in N(v_0, G_{0, t})}P(T_{v_0, v_1}^t=0)$ and use Jensen's inequality and Hoeffding’s lemma to bound it.\footnote{Proof available in our technical report \cite{technical}.}
\qed

Theorem~\ref{thm:subsample} gives us estimates that bound $p_{v_0}^N$ from above and below on expectation. However, due to sub-sampling we do not have access to $P(T_{v_0, v_1}^t=0)$ exactly. Thus, we recursively apply Theorem~\ref{thm:subsample} to obtain lower and upper bounds on $P(T_{v_0, v_1}^t=0)$, and then use those in the statement of the theorem. Thus, when modeling multi-hop infections, Theorem~\ref{thm:subsample} is recursively applied. Furthermore, in practice, we estimate $p_{min}$ from our observations (we estimate the maximum number of co-locations, $n_{max}$, between two individuals from the sub-sample, then estimate $p_{min}$ as $(1-p_{inf})^{n_{max}}$).

\textit{Final Algorithm}. Alg.~\ref{alg:inf_prob} depicts our final algorithm to find lower and upper bounds on $\hat{p}_u^t$, for an individual. This is obtained by calling $calc\_prob\_inf(<u>, t)$. The base case for Alg.~\ref{alg:inf_prob} is when $s$ contains all the nodes (the algorithm will not enter the for loop in that case), although a more efficient base case is to also consider the time $t$ and check whether there is a non-zero probability that adding another hop (i.e., line~\ref{alg:inf:rec}) will be able to spread the disease to $v_0$ until time $t$. This is possible because it takes $\mu_{IS}$ for every infected individual to spread the disease, and thus, if there are $k$ hops on the sequence $<v_0, ..., v_k>$, it takes at least $k\times \mu_{IS}$ for the disease to spread from $v_k$ to $v_0$.

\subsection{Generalizing the Diffusion Model}\label{sec:generalizing}
There are multiple underlying assumptions in Secs.~\ref{sec:upsampling:spreader} and~\ref{sec:upsampling:susceptible} that a diffusion model needs to satisfy for our methods to be applicable. First, initial infections need to be independent. That is, each individual needs to be infected independently with some initial probability. Although we have assumed this probability to be the same and $p_{init}$ for all individuals, the method can be easily modified to use probability $p_{init, u}$ for each individual $u$, i.e., the initial probabilities do not have to be the same for all individuals. 

Second, consider the function $f(l_u, l_v, t)$ that defines probability of transmission for two location sequences. For our algorithm in Sec.~\ref{sec:upsampling:spreader}, $f(l_u, l_v, t)$ is only used to calculate the edge weights of the contact network. Thus, the only requirement is for it to be possible to calculate the probability of transmission from time $s$ to time $t$ from $f(l_u, l_v, t)$ if $u$ is infected at time $s$. For the algorithm in Sec.~\ref{sec:upsampling:susceptible}, observe that the formulation in Alg.~\ref{alg:inf_prob} does not depend on the function $f$, but Alg.~\ref{alg:transmission_prob} is dependent on the specific transmission model. Here, $f$ needs to be a function that can be replaced with an algorithm that returns $P(T_{v_0, v_1, ..., v_k}^t=1|v_k\text{ gets infected at }t_i)$. This is also the case for $\Theta_{IS}$ and $\Theta_{IR}$. 

We note that although our framework is general enough to be able to handle different diffusion models, the efficiency of the calculation can be an obstacle for complex distributions. For instance, Alg.~\ref{alg:transmission_prob} in line~\ref{alg:trans:contact} takes advantage of the fact that only certain contacts between time $t_{IS}$ and $t_{R}$ can cause transmission, which is in turn because  $t_{IS}$ and $t_{R}$ are deterministically calculated (in lines~\ref{alg:trans:t_is} and~\ref{alg:trans:t_r}). However, modifying $\Theta_{IS}$ and $\Theta_{IR}$ will require checking the list of all contacts between $v_k$ and $v_{k-1}$ during which there a is non-zero probability that $v_k$ is $IS$ (as it is, the probability that $v_k$ becomes IS before $t_I+\mu_{IS}$ is zero).
\begin{table}
\begin{minipage}{0.49\columnwidth}
    \centering
    {\begin{tabular}{c|c}
        {Parameter} & {Value} \\\hline
        $d_{max}$  & $\sim11m$\\\hline
        $t_{min}$  & $15$ min.\\\hline
        $p$ & 0.01\\\hline
        $p_{init}$ & 0.1 \\\hline
        $\mu_{IS}$ & 5 days\\\hline
        $\mu_{R}$ & 12 days \\\hline
        $p_s$ & 0.1
    \end{tabular}}
    \caption{Parameter Setting}
    \label{tab:sim_setting}
\end{minipage}
\begin{minipage}{0.49\columnwidth}
    \centering
    {
    \begin{tabular}{c|c}
        Dataset & No. loc. signals \\\hline
        San Francisco & 48,699,847 \\\hline
        Manhattan & 53,147812 \\\hline
        Cook & 130,236,901
    \end{tabular}}
    \caption{No. location signals for Dec.}
    \label{tab:my_label}
\end{minipage}
\end{table}

\begin{figure}
\begin{minipage}{0.45\columnwidth}
    \centering
    \includegraphics[width=1.1\textwidth]{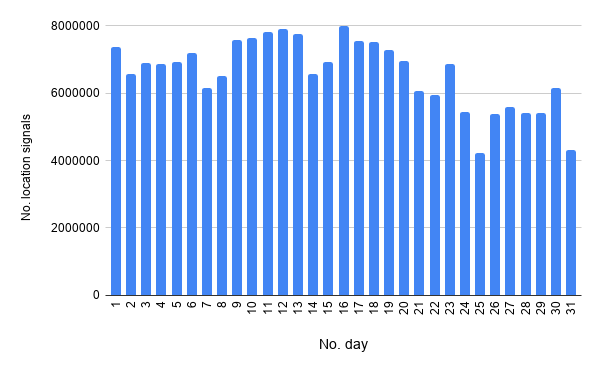}
    \caption{No. location signals per day for Manhattan dataset}
    \label{fig:LocationSignals}
\end{minipage}
\hfill
\begin{minipage}{0.45\columnwidth}
    \centering
    \includegraphics[width=\textwidth]{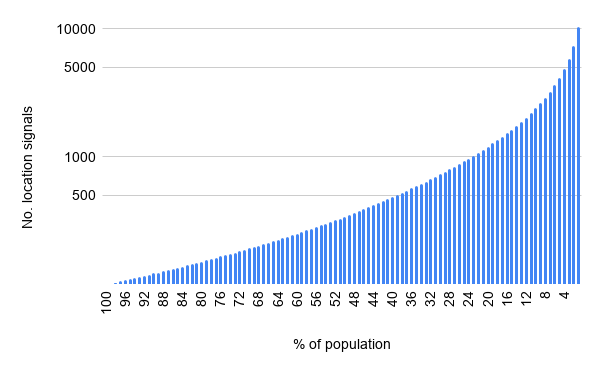}
    \caption{Distribution of location signals for Manhattan dataset}
    \label{fig:distribution}
\end{minipage}
\end{figure}

\begin{figure*}
\begin{minipage}{0.3\textwidth}
    \centering
    \includegraphics[width=\textwidth]{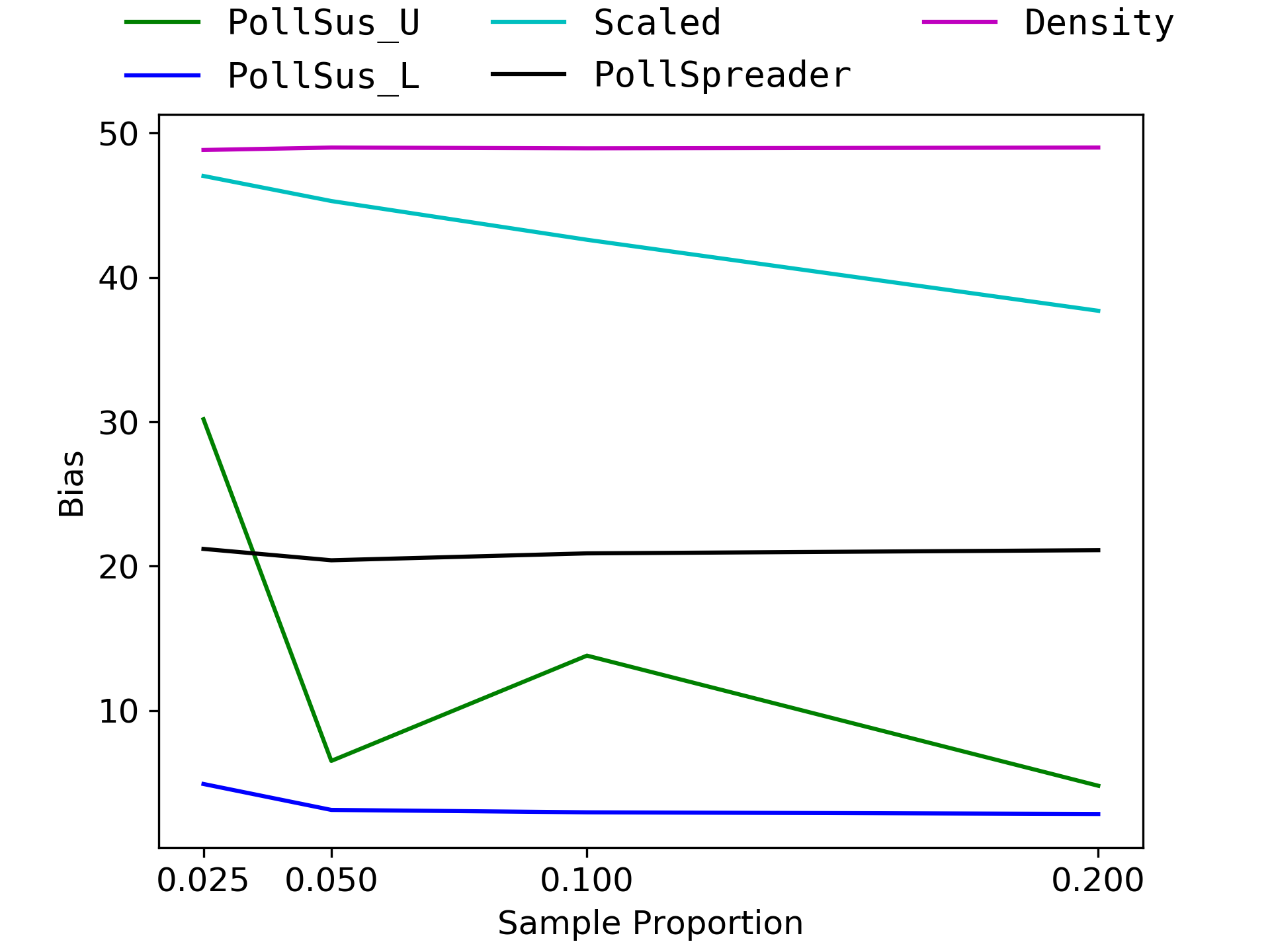}
    \caption{Bias on San Francisco dataset.}
    \label{fig:sfbias}
\end{minipage}
\hfill
\begin{minipage}{0.3\textwidth}
    \centering
    \includegraphics[width=\columnwidth]{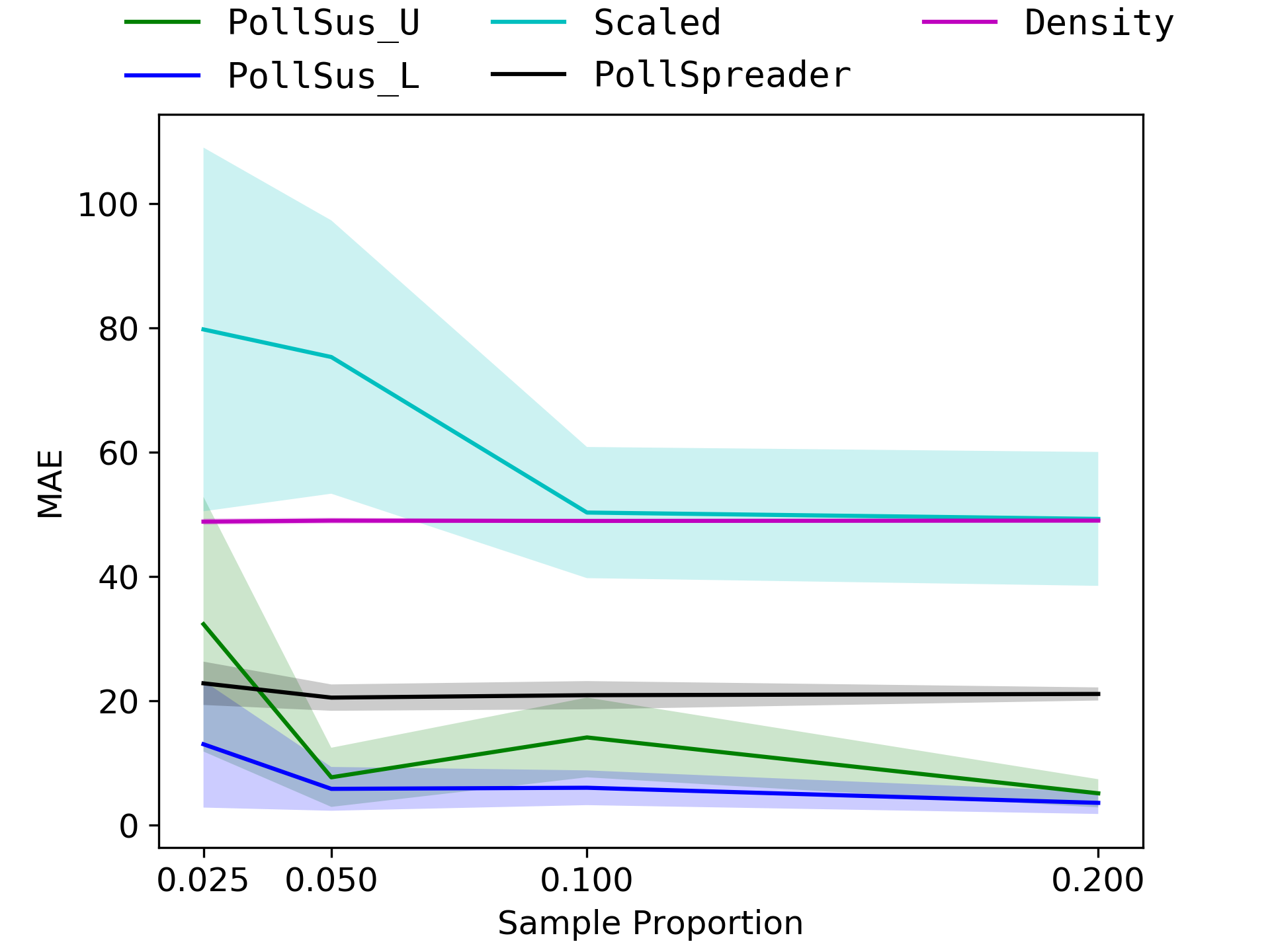}
    \caption{Error on San Francisco dataset.}
    \label{fig:sferror}
\end{minipage}
\hfill
\begin{minipage}{0.3\textwidth}
    \centering
    \includegraphics[width=\columnwidth]{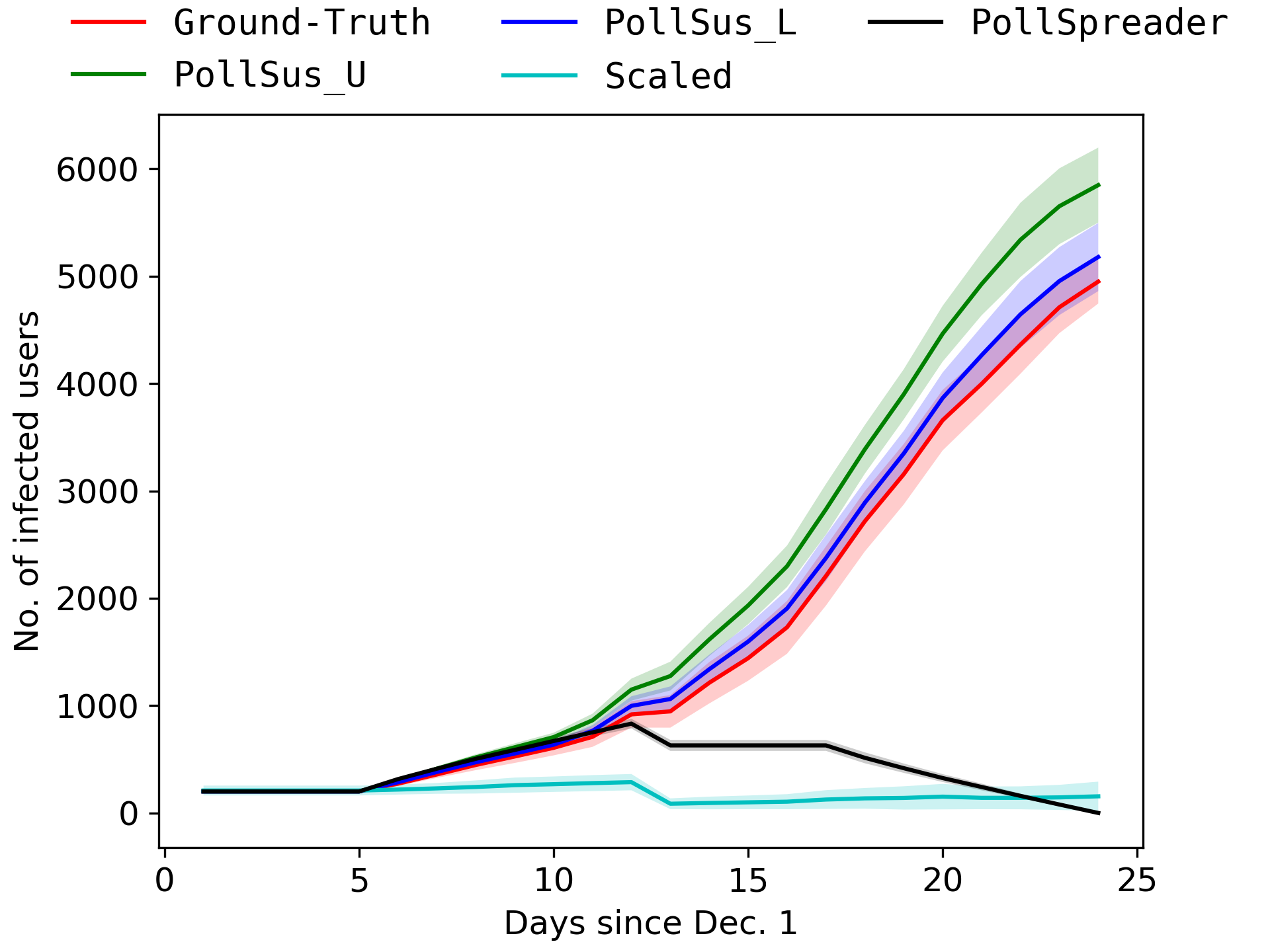}
    \caption{San Francisco infections with $p_{inf}=0.1$}
    \label{fig:sf0.1}
\end{minipage}
\vspace*{5pt}
\end{figure*}

\begin{figure*}
\begin{minipage}{0.3\textwidth}
    \centering
    \includegraphics[width=\textwidth]{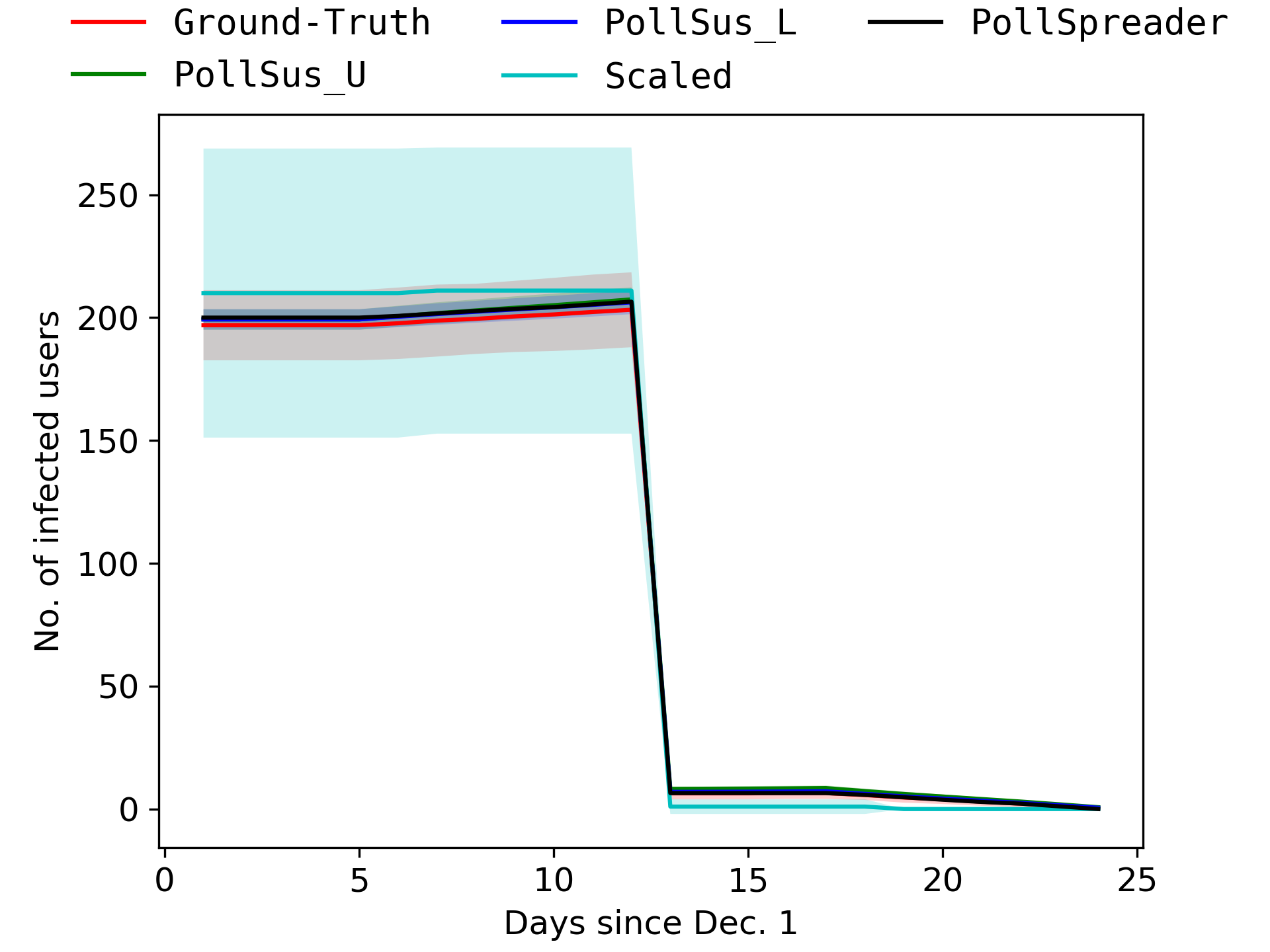}
    \caption{Cook County infections}
    \label{fig:cook0.01}
\end{minipage}
\hfill
\begin{minipage}{0.3\textwidth}
    \centering
    \includegraphics[width=\columnwidth]{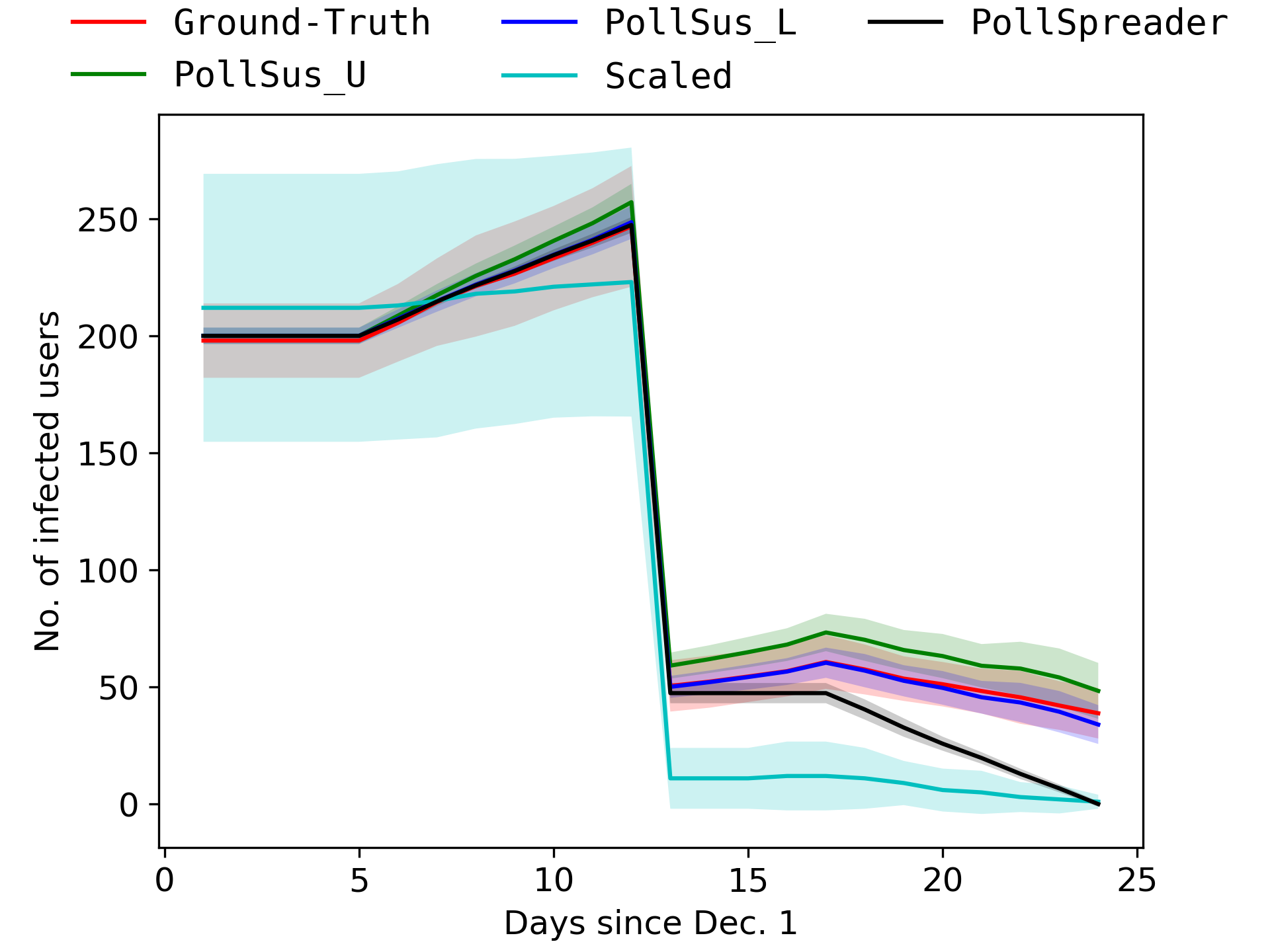}
    \caption{Manhattan infections}
    \label{fig:manhattan0.1}
\end{minipage}
\hfill
\begin{minipage}{0.3\textwidth}
    \centering
    \includegraphics[width=\textwidth]{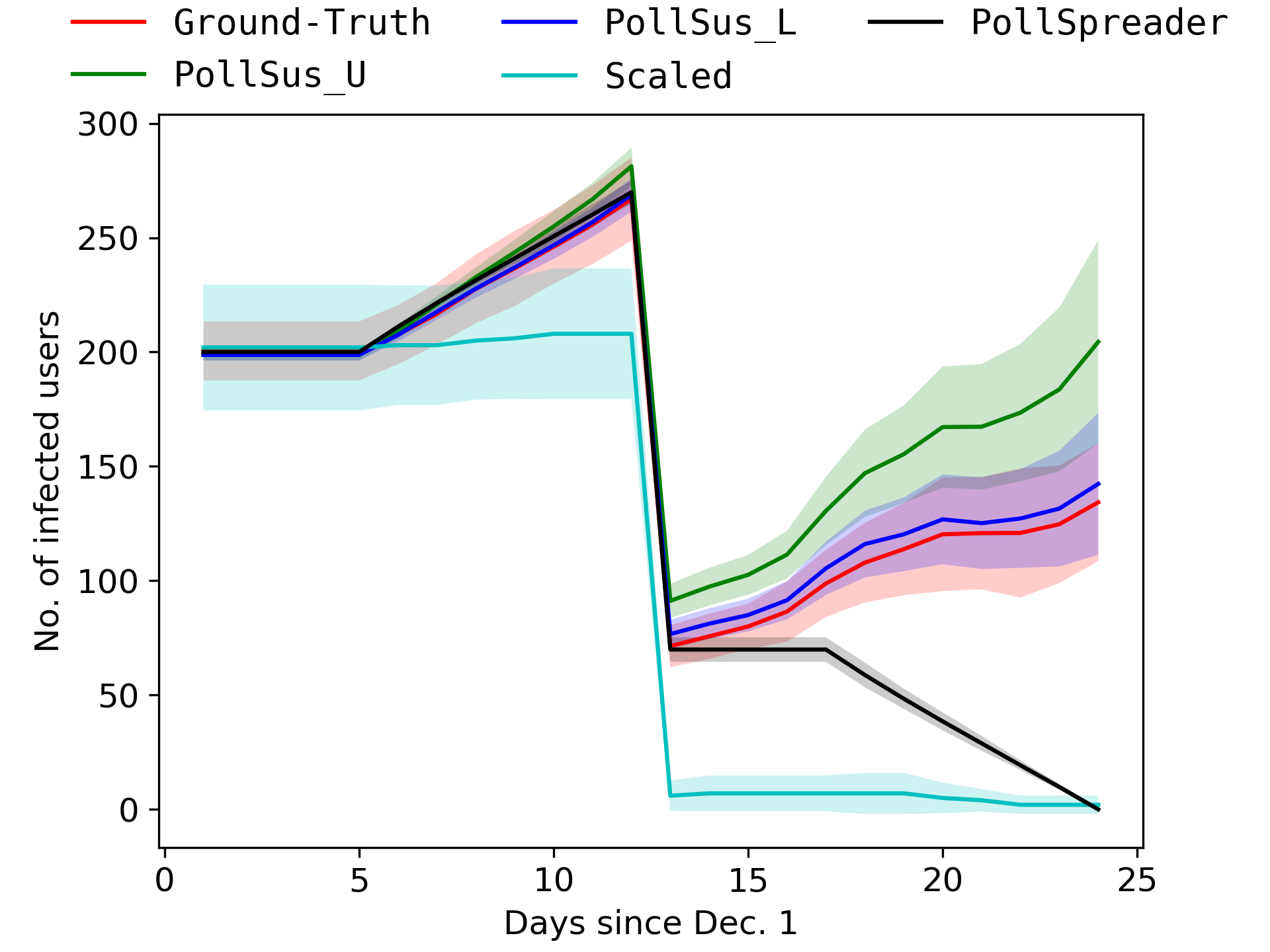}
    \caption{San Francisco infections}
    \label{fig:sf0.01}
\end{minipage}
\vspace*{5pt}
\end{figure*}

\begin{figure}
\begin{minipage}{0.45\columnwidth}
    \centering
    \includegraphics[width=1.1\textwidth]{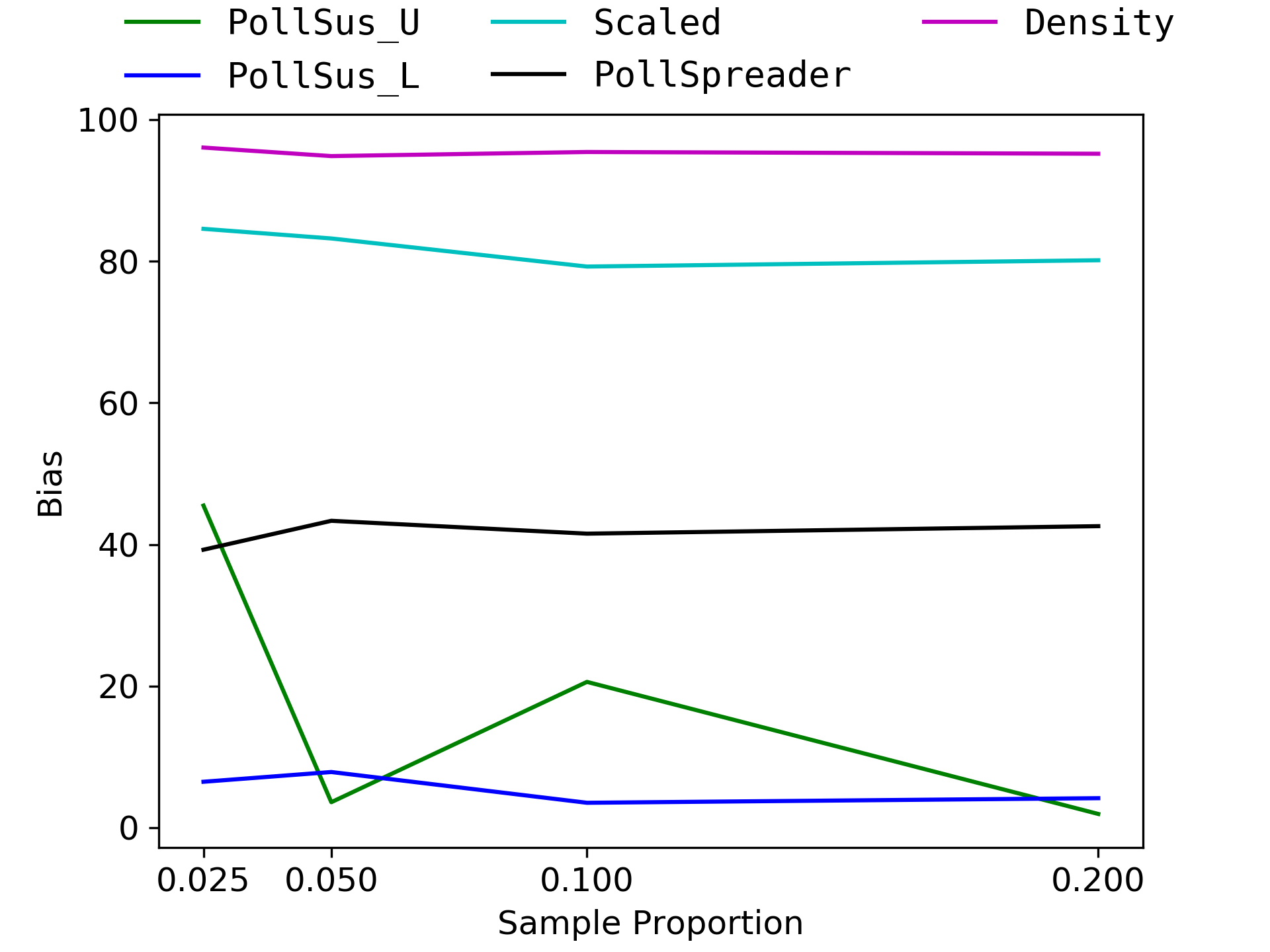}
    \caption{Bias on Gowalla dataset.}
    \label{fig:gwbias}
\end{minipage}
\hfill
\begin{minipage}{0.45\columnwidth}
    \centering
    \includegraphics[width=1.1\columnwidth]{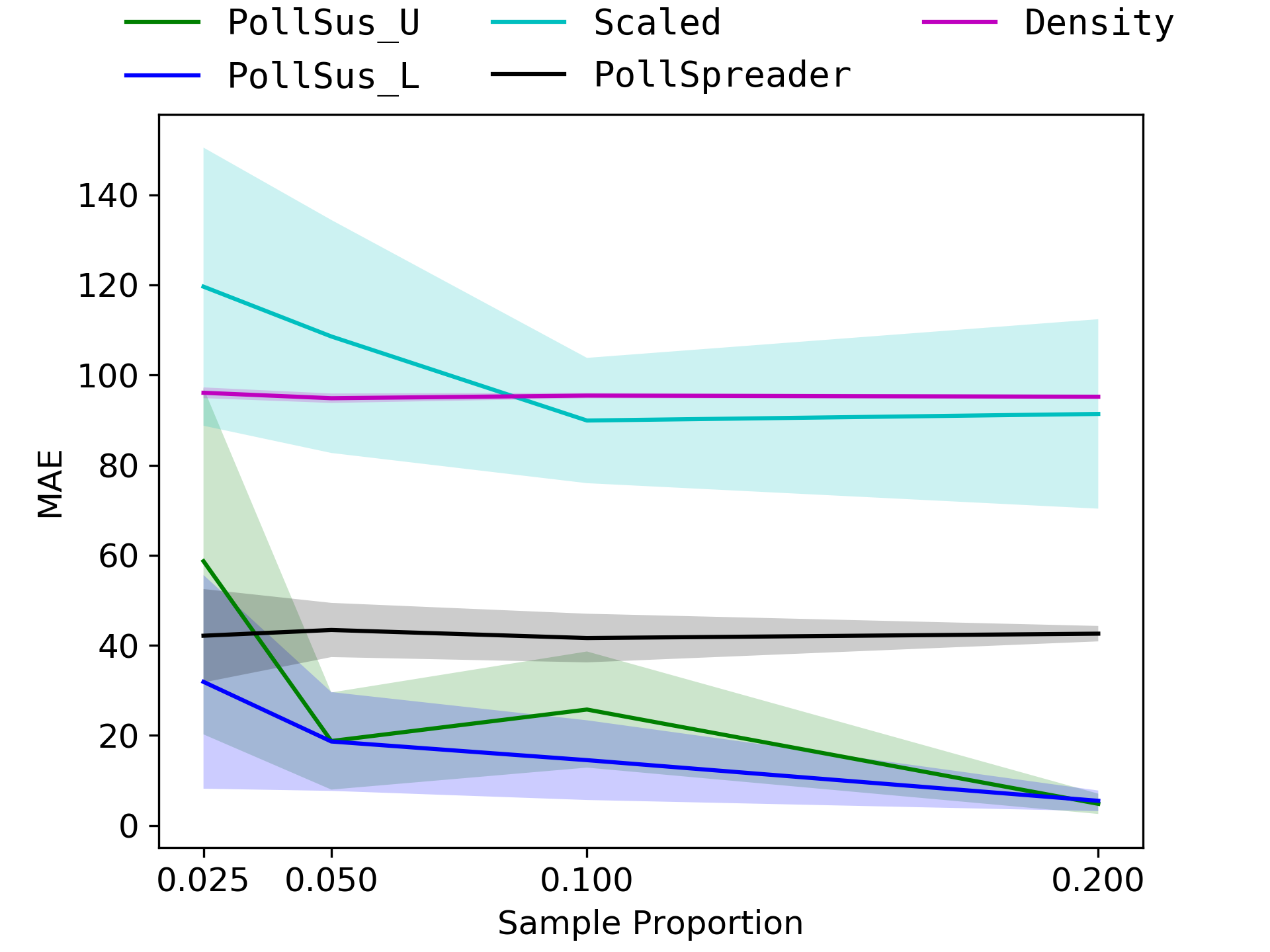}
    \caption{Error on Gowalla dataset.}
    \label{fig:gwerror}
\end{minipage}
\end{figure}

\section{Performance Evaluation}\label{sec:exp}
\subsection{Experimental Methodology}
We ran our experiments on a machine running Ubuntu 18.04 LTS equipped with an Intel i9-9980XE CPU
(3GHz) and 128GB RAM. Our experiments are designed to evaluate how close to the ground-truth our estimate is for different datasets and diffusion model parameters.

\textbf{Datasets}. We consider two datasets. Our first dataset is the Veraset dataset. Veraset \cite{veraset} is a data-as-a-service company that provides anonymized population movement data collected through location measurement signals of cell-phones across the US. We were provided access to this dataset for December 2019. The dataset consists of location signals of cell-phone devices, where each location signal is considered to be a visit, as defined in Sec. \ref{Sec:def}. Each record in the dataset consists of \texttt{anonymized}\texttt{\_device\_id}, \texttt{latitude}, \texttt{longitude}, \texttt{timestamp} and \texttt{horizontal\_accuracy}. We assume each \texttt{anonymized} \texttt{\_device\_id} corresponds to a unique individual. We discard any location signal with \texttt{horizontal\_accuracy} of worse than 25 meters. For a single day in December, there are 2,630,669,304 location signals across the US. Each location signal corresponds to an \texttt{anonymized}\texttt{\_device\_id} and there are 28,264,106 distinct \texttt{anonymized} \texttt{\_device\_id}s across the US in that day. Fig. \ref{fig:LocationSignals} shows the number of daily location signals recorded in the month of Dec. 2019 in the area of Manhattan, New York. Furthermore, Fig. \ref{fig:distribution} shows the distribution of location signals across individuals in Manhattan in Dec 2019. A point $(x, y)$ in Fig. \ref{fig:distribution} means that $x$ percent of the individuals have at least $y$ location signals in the month of Dec. in Manhattan. The Veraset dataset represents a realistic scenario where our methods can be applied, i.e., where accurate visits of a portion of the population of a city are collected and the goal is to estimate the spread over the entire population.

We use three subsets of this dataset corresponding to 20,000 individuals in San Francisco County, Manhattan County and Cook County (which contains Chicago). In the experiments, we consider this 20,000 individuals to be the entire population, based on which the ground-truth spread is calculated (Our algorithms are given access to the visits of a subset of these individuals to perform their estimation of this ground-truth spread). We used these three areas as they corresponded to different co-location patterns, which lead to differences in the spread of the disease with the same diffusion model. Specifically, for each dataset, we measured average number of daily co-locations per individual, calculated by counting the total number of co-locations in a day and dividing it by the total number of individuals. San Francisco, Manhattan and Cook counties had 2.95, 1.72 and 0.23 average number of daily co-locations per person, respectively.  

Furthermore, we use Gowalla \cite{gowalla}, a publicly available dataset to allow for reproducability of our results. Gowalla contains visits of the users obtained from a social network over several months across the US. Since the data is very sparse, we select a 20 day period with the largest number of visits and 20,000 individuals. It contains $6,760,928$ visits. 

\textbf{Algorithms}. We compare the performance of the algorithms discussed in Sec. \ref{thm:subsample}. Scale is the algorithm discussed in Sec. \ref{sec:scale}, Density was discussed in Sec. \ref{sec:density}, PollSpreader in Sec.~\ref{sec:upsampling:spreader} and PollSus\_L and PollSus\_U are the lower and upper bounds, respectively, from the method PollSusceptible discussed in Sec. \ref{sec:upsampling:susceptible}. We use Density as a surrogate for approaches that generate synthetic trajectories, e.g. \cite{feng2020learning, ouyang2018non}, where they model the probability distribution of location sequences of the individuals. However, in contrast to \cite{feng2020learning, ouyang2018non}, Density allows for directly calculating the probability of co-locations, while \cite{feng2020learning, ouyang2018non} rely on sampling to be able to find possible co-locations. For Density, we use a grid of 100$\times$100 to discretize the space (this leads to the width of each cell for Manhattan and San Francisco to be about 100m and for Cook county about 500m). 

\textbf{Metrics and Evaluation}. For each dataset, we calculate the ground-truth using Monte-Carlo simulation. This is done by simulating the spread of the disease in the entire population for whom we have data, i.e., for Veraset dataset, the 20,000 individuals. We run the simulation 10 times and take the average. To evaluate each algorithm, we sample each individual independently and with probability $p_s$, for $p_s\in \{0.025, 0.05, 0.1, 0.2\}$ to create random sub-samples of the population. For each algorithm, we measure mean absolute error (MAE) in the estimation from the ground truth. This is done by, for each set of samples, measuring the difference between the ground truth and the estimation. We also measure the \textit{bias} in each estimation. The bias is calculated by taking the average of 10 runs for each algorithm and calculating its difference with the ground truth. 

\textbf{Parameter Setting}. We use the diffusion model discussed in Sec. \ref{sec:term}, and unless otherwise stated, the parameter setting is as shown in Table \ref{tab:sim_setting}. We set the parameters of the diffusion model to mimic the spread of COVID-19. We set $\mu_{IS}$ to 5 days similar to the mean incubation period reported in \cite{li2020early}. The work in \cite{he2020temporal} reports ``Infectiousness was estimated to decline quickly within 7 days'', so we set $\mu_{R}$ to $\mu_{IS}+7$. We set $t_{min}$ to 15 min as mentioned by \cite{wiersinga2020pathophysiology}. We vary $p_{inf}\in {0.1, 0.01}$ to see how it impacts the spread, where low values of $p_{inf}$ can be seen as the case when people are wearing a mask and the larger value as when people aren't. 

\subsection{Results}
\subsubsection{Veraset Data\nopunct}\hfill\\
\noindent\textbf{Error and Bias}. Figs. \ref{fig:sfbias} and \ref{fig:sferror} show the performance of the algorithms on the San Francisco dataset. First, we observe that PollSus\_L provides very low error and bias, even when only 2.5\% of the population is sampled, and thus can be used to estimate the spread accurately. PollSus\_U performs well for higher sampling rates while Scale and Density perform poorly as they fail to adjust for infections that happen from unobserved individuals. We note that Scale also has a higher variance compared with the other methods, which is a result of the randomness of the Monte-Carlo simulation (the other algorithms are deterministic given a sub-sample). 

\noindent\textbf{Daily Infected Numbers across Counties}. Figs.~\ref{fig:cook0.01}, \ref{fig:manhattan0.1} and \ref{fig:sf0.01} show the number of infected people per day for Cook, Manhattan and San Francisco counties respectively (for clarity, we excluded Density from this experiment due to its poor performance, as shown in previous experiments) and the shaded area shows one standard deviation above and below the mean. The goal of this experiment is to examine, for different spread patterns, how closely the estimations follow the ground truth. We note that the drop in the number of infections on day 13 for all the figures is due to our choice of $\mu_{R}=12$, since all the 200 hundred initial infections recover by then.

Overall, the spread of the disease is correlated with the average number of daily co-locations per individuals, where the spread of the disease ends after the initial infections in Cook county, but it lasts longer in Manhattan and in San Francisco. We observe the idea of \textit{herd-immunity} in Fig. \ref{fig:cook0.01}, where the virus stops spreading well before all the people in the population are infected. The pattern is different for Manhattan and San Francisco, with the number of cases in Manhattan remaining at a steady level, while increasing in San Francisco (this is interesting from an epidemiology perspective, because it shows that herd immunity is dependent on the co-location pattern for a population, given the same disease). We also see that PollSus\_L and PollSus\_U follow the ground-truth closely in all the datasets and throughout the studied period. PollSpreader follows ground-truth but mainly up-to day 10, as after that  an adjustment for multi-hop infections is required which PollSpreader does not take into account. Scale accurately estimates the initial infections, as discussed in Sec. \ref{sec:scale}, but does not account for its false negatives, which gradually deteriorate its performance.

We note that, in our experiments, we have assumed the 20,000 individuals to be the true population. However, the \textit{real-world} population of each of the counties is larger (20,000 is about 0.4\% of the real-world population of Cook county and about 2.5\% of the real-world population of San Francisco). As a result the spread shown does not necessarily follow the real-world number of infections. Furthermore, since the real-world population of the counties differ, a relative comparison across the counties for the real-world number of infections is also not justified, as the up-sampling procedures would be different. Thus, we make no statement on the real-world spread of the disease. However, our method can, in fact, be used to calculate the real-world spread for each of the counties by considering the 20,000 individuals as the sub-sample of the real-world mobility pattern. We did not consider this scenario in our experiments since we did not have access to an accurate ground truth (because, first, the number of reported cases in the real-world is inaccurate \cite{infection_count} and, second, our data is for Dec. 2019).

\noindent\textbf{Daily Infected Numbers varying $p_{inf}$}. Fig.~\ref{fig:sf0.1} shows the impact of increasing $p_{inf}$. Comparing Figs.~\ref{fig:sf0.01} and \ref{fig:sf0.1} shows how increasing the probability of transmission increases the spread of the disease. Overall, the general trends are the same as before, with PollSus\_L and PollSus\_U closely following Ground-Truth. Moreover, in Fig. \ref{fig:sf0.1}, pollSpreader shows a herd immunity pattern, where the infections initially increase and then start decreasing after a certain point in time, while the true number of infections is, in fact, increasing. This is also true for Scale, where the underestimation is amplified over time, i.e., underestimating the number of current infections leads to further underestimating the people who get infected in the future. This shows the importance of an accurate estimation method, as otherwise we may observe an estimation with an entirely incorrect pattern.

\subsubsection{Gowalla Data\nopunct}\hfill\\
For our experiments on Gowalla, we set $d_{max}$ to about 110m and $p_{inf}=0.1$, since otherwise we observed very few colocations and hence infections. Figs. \ref{fig:gwbias} and \ref{fig:gwerror} show the results on the Gowalla dataset. The trends are similar to the Veraset dataset with PollSus\_L and PollSus\_U performing the best.

\section{Related Work}\label{sec:rel_work}
Studies related to our work fall into three categories: work on (1) synthetic trajectory creation, (2) contact matrices and synthetic co-location creation and (3) agent-based simulation modeling the spread of a disease.

\noindent\textbf{Synthetic Trajectory Generation}. Various methods have been proposed to generate location trajectories of people \cite{feng2020learning, ouyang2018non, jiang2016timegeo, yin2017generative}. Recent work \cite{feng2020learning, ouyang2018non} use GANs \cite{yu2017seqgan, goodfellow2014generative} to model the distribution of the trajectories, which is learned from the available samples. This method can be applied to our problem setting by sampling new trajectories from the learned model and then simulating the spread of the disease in the new population (which contains synthetic users). The simulation of \cite{feng2020learning} follows this approach, but the space is discretized. An extra \textit{contact factor} is introduced (set to a fixed value) to model the rate of co-locations between people within the same cell. This is similar to our baseline, Density (see Sec. \ref{sec:density}), which discretizes the space (the contact factor in our case is $\frac{\pi d_{max}^2}{\text{cell area}}$), and such discretization is also done in \cite{ouyang2018non}. 

Overall, there are multiple problems with these approaches. (1) Locations need to be modeled accurately to within a few meters and for long periods of time. This is a challenging task, and discretizing space, done to improve the quality of the location trajectory, makes the quality of the co-locations worse, as we observed in our experiments. (2) Mobility patterns change over time and in different cities, and the approaches need to learn different models for each of them. (3) They need large number of samples for training a neural network and they provide no theoretical guarantees on their estimation. (4) Synthetic trajectories increase the data size, which in turn requires larger computational power to estimate the spread. Our approach circumvents issue (1) by directly looking at co-locations, (2) can be directly applied to any time and location, (3) provides theoretical guarantees on its estimates and (4) performs estimation on the sub-sampled population which requires less computational power.

\noindent\textbf{Synthetic Contact Generation}. Contact matrices \cite{prem2020projecting, prem2017projecting, mossong2008social} are commonly used to simulate contact between individuals in a population. They provide aggregate level (e.g., for an entire country) contact information between different compartments in the population (e.g., rate of contacts between people of different ages) and they are estimated through surveys and diaries. However, they do not change with time and are not available for specific cities or areas, which limits their usefulness for studying the spread at a particular time and in a specific area. They furthermore do not take into account the differences in individual mobility patterns, and consider the population as multiple monoliths. Our work addresses the above issue by using location sequences of the individuals for specific periods of time and in a specific area. Finally, although more sophisticated synthetic graph generation methods, e.g., \cite{zhou2020data}, exist, we are unaware of any that do this for \textit{physical contacts}. That is, there is an extra constraint that the generated graph should correspond to physical co-locations, and be accurate for multiple weeks. We also note that such an approach would still not solve issues (2) and (3) mentioned in the previous paragraph.

\noindent\textbf{Agent-Based Simulations}. Agent-based simulations \cite{kerr2020covasim, chang2020modelling, halloran2008modeling, ferguson2020report, ferguson2006strategies} exhibit a use-case of our methodology, where spread of a disease is studied under different diffusion models and for different mobility patterns. Such simulations currently rely on fixed contact matrices, which as discussed above, limit their focus and accuracy and do not allow for study of the spread at a specific time and for a particular area. Our approach, on the other hand, can be readily used to address such inaccuracies using real data. 
\section{Conclusion}\label{sec:conc}
We studied the problem of estimating the spread of a virus, or other phenomena that can be transmitted through physical contact, in a population by having access only to a subsample of the population. We observed that modeling co-locations of the individuals, and not their locations, allows for accurate estimations. To that end, we provided two methods, PollSpreader and PollSusceptible, that estimate properties of a contact network to calculate the spread in the original population. We theoretically showed that our estimates provide lower and upper bounds on the spread in the original population, and experimentally showed that they are close to the ground-truth in practice.  Future work includes using our estimates to study different intervention strategies and studying the problem by relaxing the assumption that the samples are collected independently and uniformly at random.

\end{sloppy}

\bibliographystyle{ACM-Reference-Format}
\bibliography{references}

\clearpage
\appendix

\section{Proofs}
\textit{Proof of Theorem \ref{thm:subsample}}. Observe that 

$$\Pi_{v\in N(u, G_{0, t})}P(T_{u, v}^t=0)^{S_v}=\Pi_{v\in N(u, \hat{G}_{0, t})}P(T_{u, v}^t=0)$$ 

And is the probability of $u$ not getting infected from just its neighbours that were sampled. Denote by $\bar{T}_{u, v}^t$ the event ${T}_{u, v}^t=0$ for simplicity. To show the lower bound, we rewrite 

\begin{align*}
    (\Pi_{v\in N(u, G_{0, t})}P(\bar{T}_{u, v}^t)^{S_v})^{c_u}&=e^{\log (\Pi_{v\in N(u, G_{0, t})}P(\bar{T}_{u, v}^t)^{S_v})^{c_u}}\\
    &=e^{\sum_v c_uS_v\log P(\bar{T}_{u, v}^t)}
\end{align*}

By Jensen's inequality  
\begin{align*}
 E[e^{\sum_v c_uS_v\log P(\bar{T}_{u, v}^t)}]\geq e^{\sum_v E[S_v]c_u\log P(\bar{T}_{u, v}^t)}=e^{\sum_v p_sc_u\log P(\bar{T}_{u, v}^t)}   
\end{align*}

Setting $c_l=\frac{1}{p_s}$ we get 
$$E[e^{\sum_v \frac{S_v}{p_s}\log P(\bar{T}_{u, v}^t)}]\geq e^{\sum_v \log P(\bar{T}_{u, v}^t)}= \Pi_{v\in N(u, G_{0, t})}P(\bar{T}_{u, v}^t)$$

Which proves the lower bound.

To show the upper bound, similarly consider $E[e^{\sum_v c_uS_v\log P(\bar{T}_{u, v}^t)}]$. By Hoeffding's lemma, we have 

$$E[e^{\sum_v c_uS_v\log P(\bar{T}_{u, v}^t)}]\leq e^{{\sum_v (c_up_s\log P(\bar{T}_{u, v}^t)+\frac{c_u^2\log ^2P(\bar{T}_{u, v}^t)}{8})}}$$

Furthermore, assume $p_{min}\leq P(\bar{T}_{u, v}^t)\leq 1$. Thus, $\log^2(P(\bar{T}_{u, v}^t))\leq \log P(\bar{T}_{u, v}^t)\log(p_{min})$. So,

$$e^{{\sum_v (c_up_s\log P(\bar{T}_{u, v}^t)+\frac{c_u^2\log ^2P(\bar{T}_{u, v}^t)}{8})}}\leq e^{{\sum_v \log P(\bar{T}_{u, v}^t)(c_up_s+\frac{c_u^2\log(p_{min})}{8})}}$$

Let $c_u$ be the solution to the equation $c_up_s+\frac{c_u^2\log(p_{min})}{8}=1$, which is the solution to a quadratic equation. For such a solution, we have the right hand side above equal to $\Pi_{v\in N(u, G_{0, t})}P(\bar{T}_{u, v}^t)$ as desired and a real solution exists when $p_s\geq\sqrt{\frac{-log(p_{min})}{2}}$.

\qed

\end{document}